\documentclass[a4paper,11pt]{article}

\usepackage{jheppub}
\usepackage{bm}
\usepackage{graphicx}
\usepackage{hyperref}
\usepackage{amsmath}
\usepackage{amsfonts}
\usepackage{amssymb}
\usepackage{mathtools}
\usepackage{mathrsfs}

\newcommand{\Li}[1]{\mathop{\mathrm{Li}_{#1}}}

\def\1{\hbox{{1}\kern-.25em\hbox{l}}}

\title{Correlators of vector, tensor, and scalar composite vertices of order \boldmath $O(\alpha_s^2\beta_0)$}

\author[a]{S.~V.~Mikhailov}
\author[a,b]{and N.~Volchanskiy}

\affiliation[a]{Bogoliubov Laboratory of Theoretical Physics, JINR,
                6 Joliot-Curie, 141980 Dubna, Russia}
\emailAdd{mikhs@theor.jinr.ru}
\affiliation[b]{Research Institute of Physics, Southern Federal University,\\
                Prospekt Stachki 194, 344090 Rostov-na-Donu, Russia}
\emailAdd{nikolay.volchanskiy@gmail.com}

\abstract{
We present analytical results for massless correlators of two
vector, tensor, and scalar composite vertices with the Bjorken fractions $x$ and
$y$ of order $\alpha_s^2 \beta_0$ of QCD.
The structure of these correlators $\Pi^\text{V,T,S}(x,y; p^2)$ and properties of its main elements are discussed in detail.
Special attention is paid to verifying the results and comparing them with known particular cases.
We apply the correlators to evaluate radiative corrections to the distribution
amplitudes of  light mesons within the QCD sum rules.
}

\begin{document}
\maketitle

\section{Introduction}
\label{intro}
\begin{figure}[ht]
\centering
\includegraphics{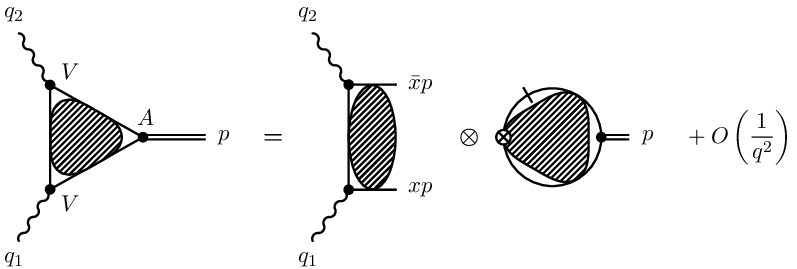}
\caption{\label{fig:VVA}Leading-twist factorization of a three-point function $\langle VVA \rangle$ into a convolution $\otimes$ of a hard four-point function and soft two-point one involving a non-local composite operator, which is denoted by the vertex \raisebox{1pt}{\tiny $\boldsymbol\otimes$}.}
\end{figure}

In this paper, we investigate massless two-point correlators of composite vertices that ``live'' on the light cone.
The local composite vertices presented below emerge in QCD due to applying the ``factorization procedure''
(or operator product expansion, OPE) to the amplitudes of hard inclusive and exclusive processes.
A well-known example of composite vertices arises from the collinear factorization of ``handbag'' diagrams in deep inelastic scattering.
 Another example related to exclusive processes is given by the $\langle V(q_1)V(q_2)A(p) \rangle$ triangle diagram
 ($V$ and $A$ are the standard vector and axial fermion currents) with hard momentum transfers $-q^2_1$, $-q_2^2 \gg p^2=(q_1+q_2)^2$.
 A two-point correlator with one composite vertex appears here as a result of factoring out $VV$ subgraphs --- the ``hard subgraphs'' of the diagram
(figure~\ref{fig:VVA}).
A correlator of two composite vertices originates from the factorization of a box diagram if we ``contract'' its hard subgraphs including the side edges
of the diagram at large values of transferred $t$.
Such a two-point correlator is a universal object that determines the asymptotic behavior of the
initial amplitude with respect to a hard momentum (i.e., in the leading twist).
Correlators like this describe the perturbative content of the hadron distribution amplitudes (DAs) --- universal hadron characteristics in the collinear approximation, which are ordered by their twist. Besides, these two-vertex correlators are important to investigate the conformal properties of composite vertices
under renormalization \cite{Craigie:1983fb}.

Let us consider some of the simplest composite bilinear fermion currents involving the $N$th derivatives of a quark field,
\begin{align}\label{eq:currents-def1}
	J_{\text{X}}^{\bar\mu}(\eta;\underline{N}) \equiv \bar{d}(\eta) \Gamma_\text{X}^{\bar\mu} \left(i\tilde{n}\nabla\right)^N u(\eta),
	\qquad
	\text{X} = \text{S, P, V, A, T},
\end{align}
where $\eta$ is a space-time point, $\nabla_\mu=\partial_\mu -i g t_a A^a_\mu$ is the covariant derivative, $\tilde{n}^\mu$ is a light-like vector, $\tilde{n}^2=0$, and $\Gamma_\text{X}^{\bar\mu}$ is a combination of the Dirac matrices, optionally carrying a string of the Lorentz indices $\bar\mu$. In particular, we are interested in the (pseudo)scalar, $\text{X}=\text{S}$ and P, vector V, axial A, and tensor T currents with, respectively,\footnote{$\hat{a} = a_\mu \gamma^\mu$, $\sigma_{\mu\nu} = (\gamma_\mu \gamma_\nu - \gamma_\nu \gamma_\mu)/2$. }
\begin{align}\label{eq:currents-def2}
	\Gamma_\text{S} = \1,
\qquad
	\Gamma_\text{P} = \gamma_5,
\qquad
	\Gamma_\text{V} = \hat{\tilde{n}},
\qquad
	\Gamma_\text{A} = \hat{\tilde{n}} \gamma_5,
\qquad
	\Gamma_\text{T}^\mu = \sigma^{\mu\nu} \tilde{n}_\nu.
\end{align}

Our goal in this work is to calculate two-point massless correlators containing the composite vertices (see \cite{Gracey:2009da}),
 e.g., the tensor-tensor $\langle T T \rangle$ correlator
\begin{equation}
i\int d^D\eta\, e^{ip\eta}
  \langle 0| \hat{\mathrm{T}} \left[J^{\mu\dagger}_{\text{T}}(\eta;\underline M)J_{\text{T}\mu}(0;\underline N)\right]|0\rangle
 = \left(\tilde{n}p\right)^{N+M+2}\Pi^\text{T}(\underline{N},\underline{M}; p^2).
\end{equation}
Further, for simplicity, we set $(\tilde{n}p)=1$.
Now, applying the inverse Mellin transforms $\hat{\text{M}}^{-1}(x\to N)$ and $\hat{\text{M}}^{-1}(y\to M)$ to $\Pi^\text{T}(\underline{N},\underline{M}; p^2)$,
one arrives at the $(x,y)$-correlator
\begin{subequations}
\begin{equation}
\hat{\text{M}}^{-1}(x \to N)\hat{\text{M}}^{-1}(y\to M)\Pi^\text{T}(\underline{N},\underline{M}; p^2) = \Pi^\text{T}(x,y;p^2) \,
\end{equation}
that depends on the longitudinal momentum fractions --- the Bjorken variables $0 \leqslant x,y \leqslant 1$ \cite{Radyushkin:1983wh}.
Here and in what follows, we underline the arguments of the images of the Mellin transform, i.e.\ our notation for the Mellin transform is
\begin{equation}
f(\underline{a}) = \hat{\text{M}}(x \to a) f(x) = \int_0^1 dx \, x^a f(x).
\end{equation}
 \end{subequations}
Note that the scalar $\langle S S \rangle$ and pseudoscalar $\langle P P \rangle$ correlators agree in the massless limit as well as a pair of axial  $\langle A A \rangle$ and vector $\langle V V \rangle$ ones:
\begin{equation}
\Pi^\text{S}(x,y;p^2) = \Pi^\text{P}(x,y;p^2),
\qquad
\Pi^\text{V}(x,y;p^2) = \Pi^\text{A}(x,y;p^2).
\end{equation}
The $(x,y)$-representation allows us to obtain \textit{any kind of composite vertices} by means of convolutions
 $\varphi(x)\otimes \Pi(x,y;p^2)\otimes \phi(y)$,\footnote{$f(x) \otimes g(x) = \int_0^1 f(x) g(x) dx$. } where the functions $\varphi$ and $\phi$ replace monomials
in the corresponding composite vertices.
Moreover, the calculation becomes much easier if we apply the inverse Mellin transforms
to the composite vertices,
$$\hat{\text{M}}^{-1}(x\to N) J_\text{X}^{\bar\mu}(\eta;\underline{N}) = J_\text{X}^{\bar\mu}(\eta;x), $$
from the very beginning \cite{Mikhailov:1984ii, Mikhailov:2018udp}.
The Feynman rules for the vertices $J_\text{X}^{\bar\mu}(\eta;x)$ are presented in appendix~\ref{app:A}.
In what follows, we will deal with the $\Pi^\text{X}(x,y;p^2)$ correlators of $x$ and $y$-vertices of different $\gamma$-matrix structures, $\text{X}=\text{S}~(\text{P})$, $\text{V}~(\text{A})$, $\text{T}$.
The key technical element necessary for our calculation --- the ``kite'' two-loop scalar integral --- was evaluated in
\cite{Mikhailov:2018udp}. In the calculation, we use the BPHZ $R$-operation in the $\overline{\rm MS}$ renormalization scheme (for dimensional regularization with $D=4-2\varepsilon$).

Along with $\Pi^\text{X}(x,y;p^2)$, we consider its Mellin moments
\begin{eqnarray} \label{def: moments-ab}
\Pi^\text{X}(x,\underline{b};p^2)=\int_0^1\Pi^\text{X}(x,y;p^2)y^b\,dy,
\qquad
\Pi^\text{X}(\underline{a},\underline{b};p^2)=\int_0^1\Pi^\text{X}(x,y;p^2)y^b x^a\, dy \,dx,
\end{eqnarray}
which are important for various applications.

The correlators calculated in this work are also important as perturbative ingredients in evaluating meson DAs within the QCD sum rule (QCD SR) approach. In this approach, the correlators are usually Borel transformed, which implies that only terms containing logarithms of $p^2$, external momentum squared, contribute to QCD SR, while the finite parts of the correlators do not survive the Borel transform. Hence, in this paper, we are mostly interested in the log-parts of the correlators.

The paper is organized as follows.
In section~\ref{sec:2}, we discuss the results of 2-loop calculations for the correlators
$\langle VV \rangle$, $\langle T T \rangle$, and $\langle S S \rangle$. We consider some checks on these results as well as their relation to the perturbative content of the corresponding DAs.
The log-part of the results has a direct physical meaning, while more lengthy nonlogarithmic parts are less interesting in the scope of this paper, see the discussion in \cite{Chetyrkin:2010dx}, and are reserved for appendix~\ref{App:B} and  \texttt{.m} files appended to the \texttt{arXiv} submission.
In section~\ref{sec:3}, we present the 3-loop expressions for the same correlators of order $O(\beta_0 a^2_s)$.
We discuss their general structure in detail and pay attention to checking their correctness. To this end, we extract some special cases of the Mellin moments (\ref{def: moments-ab}) from the results of \cite{Gracey:2009da} and compare them with the ones obtained by us.
As an immediate application, we use $\Pi^\text{X}(x,y;p^2)$ to estimate the impact of radiative corrections on different
 meson distribution amplitudes in section~\ref{sec:4}.
For all cases, the radiative contributions to the DAs look significant and should be taken into account in future estimations.
In section~\ref{sec: concl}, we formulate  our conclusions. Some important technical details and part of the results are given in five appendices.


\section{Correlators \boldmath $\langle V\,V \rangle$, $\langle T\,T \rangle$, $\langle S\,S \rangle$ at NLO}
 \label{sec:2}

\begin{figure}[ht]
\centering
\includegraphics[width=.24\linewidth]{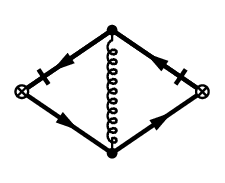}
\includegraphics[width=.24\linewidth]{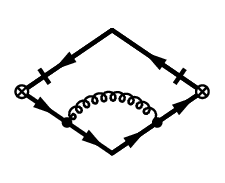}
\includegraphics[width=.24\linewidth]{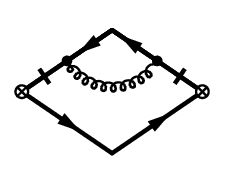}
\includegraphics[width=.24\linewidth]{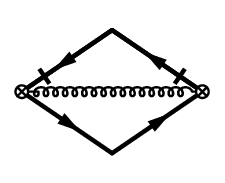}

\includegraphics[width=.24\linewidth]{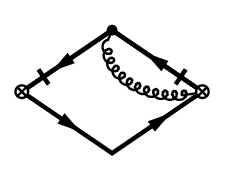}
\includegraphics[width=.24\linewidth]{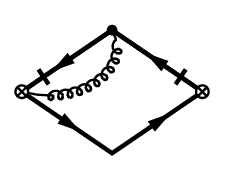}
\includegraphics[width=.24\linewidth]{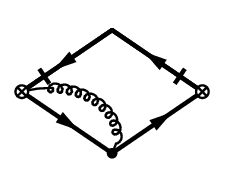}
\includegraphics[width=.24\linewidth]{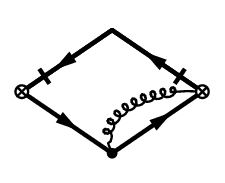}

\caption{\label{fig:NLOdiags}Feynman diagrams for a two-point function with two composite vertices of order $a_s$.}
\end{figure}

In pQCD, the $p^2$-dependence of the correlators manifests itself through the logarithm $ L=\ln\left(-p^2/\mu^2\right)=\ln\left(P^2/\mu^2\right) $, except for the case of $\langle S\,S \rangle$ ($\langle P\,P \rangle$) containing, also, a common factor of $P^2 = -p^2$ (see the definition in section~\ref{sec:SS-PP-NLO}):
\begin{subequations}
\begin{eqnarray}
\Pi^\text{X}(x,y; P^2)&=& \sum_{i=0}a_s^{i}(\mu^2) \sum_{j=0}^{i+1}\Pi_{i,j}^\text{X}(x,y)~L^{j}=\sum_{i=0}a_s^{i}(\mu^2) \Pi^\text{X}_i(x,y; P^2), \label{eq:Pi^X}\\
    && \Pi^\text{X}_i(x,y; P^2)=\sum_{j=0}^{i+1}\Pi^\text{X}_{i,j}(x,y)~L^{j},
    \qquad
    a_s = \frac{\alpha_s}{4\pi}\,. \label{eq:Pi_i}
\end{eqnarray}
\end{subequations}
The generalized one-loop ERBL evolution kernels $a_s C_F V_\text{X}$ are important and natural elements in the calculations of the corresponding $\Pi_i^\text{X}$ \cite{Mikhailov:1988nz-JINRrep}.
These kernels are generated by all subgraphs with a composite vertex that are contracted to be substituted by counterterms as required by the BPHZ $R$-operation.
 Therefore, in our results, all the leading-log terms $\Pi_{i,i+1}^\text{X}$, counterterm contributions, and some other parts of the correlators are proportional to the kernels and their generalizations, see below.
We shall start with the vector-vector correlator and the corresponding $V_\text{V}$ kernel.

\subsection{\boldmath $\langle V\,V \rangle(x,y)$ correlator}

Evaluating the correlator $\Pi^\text{V(A)}(x,y;p^2)$,
\begin{equation}
 (\tilde{n}p)^2~\Pi^\text{V}(x,y;p^2)= i \int d^D\eta \, e^{ip\eta} \langle 0|\hat{\mathrm{T}}\left[J_\text{V}^\dagger(\eta;x)J_\text{V}(0;y) \right]|0 \rangle,
\end{equation}
where the current $J_\text{X}(\eta;x)$ is defined by eqs.~\eqref{eq:currents-def1} and \eqref{eq:currents-def2}, it is convinient and natural to express the result in terms of some ``building blocks'' \cite{Mikhailov:1984ii}~--- the LO function $d(y;\varepsilon)$ (which is proportional to the one-loop correlator) and, starting from NLO, the generalized kernels $V_a(x,y;\varepsilon)$ and $V_b(x,y;\varepsilon)$:\footnote{The generalized kernels appear as  $V(x,y;a_s \gamma_g)$ after summing up renormalon chains in the one-loop kernels \cite{Mikhailov:1997zg,Mikhailov:1998xi}, where the infinitesimal dimensional-regularization parameter $\varepsilon$ is replaced with $a_s \gamma_g$.}
\begin{equation}
\hat{\mathbf{S}} f(x,y) = f(x,y) + f(\bar x, \bar y),
\qquad
 \hat{\mathbf{P}} f(x,y) = f(x,y) + f(y,x)\nonumber;
\end{equation}
\begin{subequations}
\label{eq:VecV}
\begin{eqnarray}
&&V_a(x,y;\varepsilon)_+= 2\left[ \hat{\mathbf{S}}\theta (y>x) \left( \frac{x}{y}\right)^{1+\varepsilon}\right]_+, \qquad V_a(x,y)\equiv V_a(x,y;0),\label{eq:VecVa}\\
&&V_b(x,y;\varepsilon)_+= 2\left[\hat{\mathbf{S}}\frac{\theta (y>x)}{y-x}\left(\frac{x }{y}\right)^{1+\varepsilon}\right]_+, \qquad V_b(x,y)\equiv V_b(x,y;0); \label{eq:VecVb}\\
&&V^{(0)}(x,y)_{+}=V^{}_{a}(x,y)_++V^{}_{b}(x,y)_+=2\left[\hat{\mathbf{S}}\theta(y>x)\frac{x}{y}\left(1+\frac{ 1}{y-x}\right)\right]_+ \,; \label{eq:VecVc} \\
&&d(y; {\varepsilon})= (y\bar{y})^{1+\varepsilon};
\qquad
d\equiv d(y)=d(y; {0}),
\qquad
\dot{d}=\left. \frac{d}{d\varepsilon}d(y;\varepsilon)\right\vert_{\varepsilon=0}\, , \label{eq:d-def}
\end{eqnarray}
\end{subequations}
where the part $V_b$ of the complete kernel absorbs the contributions with a gluon leg (or a renormalon chain) attached to the composite vertex, while $V_a$ corresponds to all other topologies contributing to the one-loop kernel. The part $V_a$ and $V_b$ of the complete kernel enter in $\Pi_{i,j}^\text{V}$ in different ways.
Here, $a_s C_{F} V^{(0)}_{+}$ is the one-loop ERBL kernel,
which describes the ERBL evolution of the DAs of the longitudinally polarized vector ($\rho$) and pseudoscalar ($\pi$) mesons (see appendix~\ref{App:DA}).
The plus-distribution form of the $V_\text{V}$ kernels is the general property for any number of loops~--- it is the consequence
of the vector (axial) current conservation,
its anomalous dimension being $\gamma(0) \sim \int_0^1dx~V(x,y)_{+}=0$. Therefore, the kernel can be written as
$$V(x,y)_{+}=V(x,y)-\delta(x-y)\int_0^1 dt \, V(t,y)\,.$$
Higher derivatives of $V_{a,b}(x,y; \varepsilon)$ and $d(y; \varepsilon )$ with respect to $\varepsilon$ proliferate in expressions for higher orders in $a_s$ \cite{Mikhailov:1985cm}.

The LO $\langle VV \rangle$ correlator can be written as
\begin{eqnarray}
\Pi^\text{V}_{0,1} &=&-\frac{N_c}{2\pi^2}d(y)\delta(x-y),
\qquad
\Pi^\text{V}_{0,0}=-\frac{N_c}{2\pi^2}\dot{d}(y)\delta(x-y)
\end{eqnarray}
in terms of the derivatives of the one-loop function $d(y;\varepsilon)$.

The NLO $\langle VV \rangle$ correlator (figure~\ref{fig:NLOdiags}) obtained in an arbitrary covariant gauge reads
\begin{subequations}
 \label{eq:PiV1}
\begin{eqnarray}
 \Pi^\text{V}_1 &=& \frac{N_c}{\pi^2} C_F \sum_{j=0}^{2} \tilde\Pi^\text{V}_{1,j}(x,y) L^j;
\end{eqnarray}
\begin{eqnarray}
\tilde{\Pi}^\text{V}_{1,2}&=& \frac12 W^{(0)}_{+}= \frac12 V^{(0)}_+ d,  \label{eq:PiV12}\\
\tilde{\Pi}^\text{V}_{1,1}&=& \left( \dot{W}_a + \dot{W}_b - W_a \dot{h}_a - W_b \dot{h}_b \right)_+ - \dot{H}_a W_a
- \frac12 \hat{\mathbf{P}} \left[ V^{(0)}_{+(x)}~ \dot{d} \right],
 \label{eq:PiV11}
\end{eqnarray}
 \end{subequations}
where $d=d(y)$, $\dot{d}=\dot{d}(y)$, and all other quantities are functions of $x$ and $y$, i.e.\ $V=V(x,y)$, $W_a =W_a(x,y)$, etc. The functions $W_i=W_i(x,y)$, $i=a,\, b$ with dots are the coefficients of the Taylor expansion for the function $W_i(x,y; \varepsilon) =  V_i(x,y; \varepsilon) d(y; \varepsilon)$, i.e.\ $W_i=V_i d$ and $\dot{W}_i = \left. \frac{d}{d\varepsilon} W_i(x,y ; {\varepsilon}) \right\vert_{\varepsilon=0}$. The quantities $\dot{H}_a,~\dot{h}_a, ~\dot{h}_b$ are the symmetric functions presented in appendix~\ref{App:B} together with the nonlogarithmic term $\tilde{\Pi}^\text{V}_{1,0}$ (\ref{eq:PiV10}), which, as far as we know, has never been calculated before. The plus distributions for a function $f(x,y)$ are defined as
\begin{subequations}
\begin{align}
	f(x,y)_{+} \equiv f(x,y)_{+(x)} &{}= f(x,y) - \delta(x-y) \int_0^1 dt\, f(t,y),
	\\
	f(x,y)_{+(y)} &{}= f(x,y) - \delta(x-y) \int_0^1 dt\, f(x,t).
\end{align}
\end{subequations}
The expressions in eqs.~(\ref{eq:PiV12}) and (\ref{eq:PiV11}) coincide with the ones obtained in \cite{Mikhailov:1988nz-JINRrep}.
The 0th moment $\Pi^\text{V}_{1,1}(x,\underline{0})$ was evaluated in \cite{Ball:1996tb,Mikhailov:1988nz-JINRrep,Mikhailov:1988nz} and a few first two-fold Mellin moments of the complete correlator $\Pi^\text{V}_{1(2)}(\underline{N},\underline{M}) $ were computed in \cite{Gracey:2009da}. We will come back to that in section~\ref{sec:VVcorr-check} to verify our results.

Let us mention important features of the coefficient $\Pi_{n,n+1}^\text{V}(x,y)$, in particular the leading-log NLO term in (\ref{eq:PiV1}):
\begin{enumerate}
\item The leading-log term $\tilde{\Pi}^\text{V}_{1,2} \sim W^{(0)}_+$ can be diagonalized by the
``standard'' Gegenbauer polynomials
$\{C^{(3/2)}_k(y-\bar{y})\}$ of the index $3/2$, while the other terms, $\tilde{\Pi}^\text{V}_{1,1}$ and $\tilde{\Pi}^\text{V}_{1,0}$, cannot be diagonalized in this way.

\item Due to the vector-current conservation, the one-fold 0th moments of the leading-log terms vanish,
\begin{equation}
\Pi^\text{V}_{1,2}(x) \equiv \Pi^\text{V}_{1,2}(x,\underline{0}) = \int_0^1 dy \, \Pi^\text{V}_{1,2}(x,y)=0 . \label{eq:currveccons}
\end{equation}
 \end{enumerate}
It should be stressed that the identity  $\Pi_{n,n+1}^\text{V}(x,\underline{0})=0$ originates from the vector-current conservation and ($x \leftrightarrow y$) permutation symmetry rather than particular properties of a specific calculation; therefore, it holds true not only in NLO, but in any higher loop orders as well.

The zeroth moment of the correlator,
\begin{equation}
\Pi^\text{V}(x;P^2) = \Pi^\text{V}(x,\underline{0};P^2) = \int_0^1dy~ \Pi^\text{V}(x,y;P^2),
\end{equation}
is the source of perturbative contributions to the QCD sum rules for the meson DAs $\varphi^{}_{M}$ with appropriate meson quantum numbers \cite{Mikhailov:1988nz-JINRrep}.
We will discuss it in more details at the beginning of section~\ref{sec:4} and mention here only that the Borel transformed correlator $\Pi^\text{V}(x;P^2)$ determines $\Delta \varphi_{M_{\parallel}}$~--- perturbative part of the DA $\varphi^{}_{M_{\parallel}}$ for the leading twist of $\pi$ mesons and longitudinally polarized vector mesons such as $\rho_{\parallel}$.
Indeed,  applying the Borel transform $\hat{\mathbf{B}}_{(M^2)}$ to $\Pi^\text{V}(x;P^2)$,\footnote{Here and below, $\hat{\mathbf{B}}_{(M^2)}$ stands for the Borel transform with respect to $P^2$, $P^2 \to M^2$.
The definition and special cases necessary to deal with the correlators of this paper are given in section~\ref{sec:4} and appendix~\ref{App:E}.}
we arrive at the well-known NLO expression \cite{Mikhailov:1988nz}
\begin{align} \label{eq:NLOcorr}
\Delta \varphi^{(0+1)}_{M_{\parallel}}(x;M^2) =
     \hat{\mathbf{B}}_{(M^2)}\Pi^\text{V(A)}_{\text{0+1}}(x,P^2)= \frac{ N_c}{2\pi^2} x \bar{x} \left\{ 1+ a_s C_F \left[5-\frac{\pi^2}{3}+\ln^2\left(\frac{\bar{x}}{x}\right)\right]\right\},
\end{align}
where $M^2$ is the Borel parameter.
The radiative content of the $\pi$ and $\rho_\parallel$ meson DAs of twist-2 will be considered further in section~\ref{sec:rhoL}.

\subsection{\boldmath $\langle T\,T \rangle(x,y)$ correlator}
Let us recall the definition of the tensor-tensor correlator,
\begin{equation}
 (\tilde{n}p)^2~\Pi^\text{T}(x,y;p^2)= i \int d^D\eta \, e^{ip\eta} \langle 0|\hat{\mathrm{T}}\left[J^{\dagger}_{\text{T}\mu}(\eta;x)J^\mu_{\text{T}}(0;y)\right]|0 \rangle,
\end{equation}
 and the components of the corresponding one-loop  ERBL kernels \cite{Mikhailov:2008my}, $V_\text{T}=a_s C_F V^{(0)}_\text{T}$,
\begin{eqnarray}
 V^{(0)}_\text{T}= V_{b}(x,y)_+ -\delta(y-x), \qquad V_a^\text{T}= -\delta(y-x), \qquad V_{b}^\text{T} = V_b(x,y;0). \label{eq:TVab}
\end{eqnarray}
As in the case of one-loop vector kernel, $V^\text{T}_b$ designates contributions from the composite vertices with a gluon leg, while $V^\text{T}_a$ correspond to all others. In the tensor case, however, the part $V^\text{T}_a$ comes solely from the quark-propagator radiative corrections, which makes it trivially ``diagonal''. We write it explicitly in what follows.
The $\langle TT \rangle$ correlator at NLO can be written in terms of the tensor kernels, the derivative $\dot{W}_b$ introduced in the previous subsection, and the one-loop functions $d$ and $\dot{d}$ of eq.~\eqref{eq:d-def}:
\begin{eqnarray}
\Pi^\text{T}_{0,1} &=&\frac{N_c}{\pi^2}d(y)\delta(x-y), \qquad \Pi^\text{T}_{0,0}=\frac{N_c}{\pi^2}\left[ \dot{d}(y) + d(y) \right] \delta(x-y);
\end{eqnarray}

\begin{subequations}
\label{eq:PiT1}
\begin{eqnarray}
 \Pi^\text{T}_1 &=& -2 \frac{N_c}{\pi^2} C_F \sum_{j=0}^{2} \tilde\Pi^\text{T}_{1,j}(x,y) L^j;\\
\tilde{\Pi}^\text{T}_{1,2}&=& \frac12 W^{(0)}_\text{T}\equiv \frac12 W_{b+} - \frac12 y \bar{y} \delta(x-y),  \label{eq:PiT12}\\
\tilde{\Pi}^\text{T}_{1,1}&=& \mathop{\hat{\mathbf{S}}} \left[ \theta(\bar{z} > 0)  \ln ( \bar z ) \right] +  \left( \dot{W}_b + W_b \ln\lvert x-y \rvert \right)_+
- \frac12\mathop{\hat{\mathbf{P}}} \left[ \left( V_b\, \dot{d} \right)_{+(x)} \right]
\nonumber\\
&&{}+ \delta(x-y) \left(  d - \frac12 \dot{d} \right), \label{eq:PiT11}
\end{eqnarray}
where
\begin{equation}
 W^{(0)}_\text{T} = V^{(0)}_\text{T} d = W_{b+} - y \bar{y} \delta(x-y),
 \qquad
  z=(y \bar x) /(x \bar y)\,, \label{eq:W_T}
\end{equation}
\end{subequations}
and the variable $z$  is the conformal ratio \cite{Mikhailov:2018udp,Braun2017}.
The nonlogarithmic part $\tilde{\Pi}^\text{T}_{1,0}$
is presented in (\ref{eq:PiT10}) of appendix~\ref{App:B}. All the calculated parts of $\tilde{\Pi}^\text{T}_{1}$ agree with the two-fold 0th moment $\tilde{\Pi}^\text{T}_{1}(\underline{0},\underline{0})$ computed in \cite{Gracey:2009da}.

After applying the Borel transform to it, the correlator $\Pi^\text{T}(x;P^2)\equiv \Pi^\text{T}(x,\underline{0};P^2)$ constitutes the perturbative part $\Delta \varphi^{}_{M_{\perp}}$ of the twist-2 DA $\varphi^{}_{M_{\perp}}$ describing the
transversely polarized vector mesons such as the $\rho_\perp$ meson \cite{Ball:1996tb}:
\begin{eqnarray}
\Delta \varphi^{(0+1)}_{M_{\perp}}(x;M^2) & =& \hat{\mathbf{B}}_{(M^2)}\Pi^\text{T}_{\text{0+1}}(x,P^2) \nonumber\\
 & =& \frac{ N_c}{2\pi^2} x \bar{x} \left\{ 1+ a_s C_F \left[6-\frac{\pi^2}{3}+\ln^2\left(\frac{\bar{x}}{x}\right)+\ln(x\bar{x})+2L_\text{B}\right]\right\}\,. \label{eq:TTNLO}
\end{eqnarray}
 The $\Delta \varphi^{(0+1)}_{M_{\perp}}$ depends on the logarithm of the Borel parameter, $L_\text{B}= \ln\left(\frac{M^2}{\mu^2}e^{-\gamma_\text{E}}\right)$, since
 the tensor current is not conserved,
 see $\tilde{\Pi}^\text{T}_{1,2}$ in (\ref{eq:PiT12}).
 The above expression for $\Delta \varphi^{(0+1)}_{M_{\perp}}$ was first derived in \cite{Ball:1996tb}.

\subsection{\boldmath $\langle S\,S \rangle(x,y)$ correlator}\label{sec:SS-PP-NLO}

The scalar-scalar correlator is defined as
\begin{gather}
p^2 \Pi^\text{S}(x,y;p^2)= i \int d^D\eta \, e^{ip\eta} \langle 0 \lvert \hat{\mathrm{T}}\left[J_\text{S}^\dagger(\eta;x)J_\text{S}(0;y)\right] \rvert 0 \rangle ,
\end{gather}
and the components of the ERBL one-loop kernel corresponding to the scalar composite vertex are
\begin{eqnarray}
V^\text{S}_a(x,y) = 2\left(\hat{\mathbf{S}} \frac{\theta (y>x)}{y} \right)_+ +3\delta(y-x),
\quad
V^\text{S}_b(x,y) = 2\left[\hat{\mathbf{S}} \left( \frac{x}{y}\frac{\theta(y>x)}{y-x}\right) \right]_+. \label{eq:scalarVab}
\end{eqnarray}
In contrast to the vector kernel, the total scalar ERBL kernel $V_\text{S}= a_s C_F V^{(0)}_\text{S}$,
\begin{eqnarray}
V^{(0)}_\text{S}(x,y)=V^\text{S}_a(x,y)+V^\text{S}_b(x,y)=2\left(\hat{\mathbf{S}}\frac{\theta (y>x) }{y-x}\right)_+ +3 \delta(y-x), \label{eq:scalarV}
\end{eqnarray}
 is already symmetric by itself, $V^{(0)}_\text{S}(x,y)=V^{(0)}_\text{S}(y,x)$. It is diagonalized  in the basis of the Gegenbauer polynomials $C^{(1/2)}_n(y-\bar{y})$.
The eigenfunctions and corresponding eigenvalues of $V^{(0)}_\text{S}$ are $\{C^{(1/2)}_n(y-\bar{y}), \gamma_n \}$, where $\gamma_n=3-4\left[\psi(n+1)-\psi(1)\right]$.

The one-loop scalar-scalar correlator (prior to expanding it in $\varepsilon$) is proportional to the function
\begin{equation}
d_\text{S}(y;\varepsilon) = (y\bar{y})^{\varepsilon},
\end{equation}
its first Taylor coefficients being
\begin{equation}
d_\text{S}\equiv d_\text{S}(y; 0) =1,
\qquad
\dot{d}_\text{S}=\left. \frac{d}{d\varepsilon}d_\text{S}(y;\varepsilon)\right\rvert_{\varepsilon=0}=\ln(\bar{y}y),
\qquad
\ddot{d}_\text{S}=\left. \frac{d^2}{d\varepsilon^2}d_\text{S}(y;\varepsilon)\right\rvert_{\varepsilon=0}=\ln^2(\bar{y}y). \label{eq:dS}
\end{equation}

The components of the expansion \eqref{eq:Pi^X} for the correlator $\Pi^\text{S}(x,y;p^2)$ can be naturally expressed
using the functions in eqs.~(\ref{eq:scalarVab})--\eqref{eq:dS}:
\begin{eqnarray}
\Pi^\text{S}_{0,1} =-\frac{N_c}{8\pi^2}d_\text{S}(y)\delta(x-y),
\qquad
\Pi^\text{S}_{0,0}=-\frac{N_c}{8\pi^2}\dot{d}_\text{S}(y)\delta(x-y)\,, \label{eq:PiS0}
\end{eqnarray}
\begin{subequations}
 \label{eq:PiS1}
 \begin{eqnarray}
 \Pi^\text{S}_1 &=& \frac{N_c}{8\pi^2} C_F \sum_{j=0}^{2} \tilde\Pi^\text{S}_{1,j}(x,y) L^j; \\
\tilde{\Pi}^\text{S}_{1,2}&=& V^{(0)}_\text{S},  \label{eq:PiS12}\\
\tilde{\Pi}^\text{S}_{1,1}&=& 2 \hat{\mathbf{P}} \left\{ \left[ \hat{\mathbf{S}} \frac{\theta(y>x)}{y-x}\left(\ln(y-x)+\ln\left(\frac{x}{y}\right) - \frac{x}{y} \right) + \frac12 \right]_{+(x)} \right\}
\notag\\
&& {}+ \delta(y-x)\left(3 \dot{d}_\text{S} - 11 \right), \label{eq:PiS11}
\end{eqnarray}
\end{subequations}
The moments $\tilde{\Pi}^\text{S}_{i,j}(\underline{0},\underline{0})$ for all the terms in eq.~(\ref{eq:PiS1}) coincide with the results in \cite{Gracey:2009da}.
In contrast to $\Pi^\text{V}$ and $\Pi^\text{T}$ cases,
the correlator $\Pi^\text{S}(x;P^2)\equiv \Pi^\text{S}(x,\underline{0};P^2)$ might be related to the pion DA of twist 3, $\varphi^p_{3;\pi}(x)$,
see appendix~\ref{App:DA} and, e.g.\ \cite{Ball:1998je,Ball:2006wn}.
Below, we present $ \hat{\mathbf{B}}_{(M^2)} \left[ P^2 \Pi^\text{S}_{0+1}(x;P^2) \right]$~--- a possible source of perturbative contribution $\Delta \varphi^p_{3;\pi}$ to $\varphi^p_{3;\pi}$:
\begin{eqnarray}
\hat{\mathbf{B}}_{(M^2)} \left[-P^2 \Pi^\text{S}_{0+1}(x;P^2) \right] =\frac{N_c}{8\pi^2} M^2 \Big\{1+ a_s C_F \Big[ 5-3\ln(\bar{x}x) - 6 L_\text{B} \Big]  \Big\}.
\end{eqnarray}
This radiative correction at NLO is a new result. The origin of its non-$L_\text{B}$ piece is the trivial integral of the term with the Dirac delta function in (\ref{eq:PiS11}), while the $L_\text{B}$ part stems from (\ref{eq:PiS12}).

The nonlogarithmic part $\tilde{\Pi}^\text{S}_{1,0}$ of the two-loop correlator \eqref{eq:PiS1} vanishes under the Borel transform and is given in eq.~\eqref{eq:PiS10} of the appendices.


\section{Correlators \boldmath $\langle V\,V \rangle$, $\langle T\,T \rangle$, and $\langle S\,S \rangle$ of order $a_s^2\beta_0$}
\label{sec:3}
Let us focus on the N$^2$LO expansion in eq.~(\ref{eq:Pi_i}),
\begin{eqnarray} \label{eq:Pi_2}
 && \Pi^\text{X}_2(x,y; P^2)=\sum_{j=0}^{3}\Pi^\text{X}_{2,j}(x,y)~L^{j}\,.
\end{eqnarray}
In order $a_s^2$, the coefficients $\Pi^\text{X}_{2,3}$ at $L^3$, the highest power of $L$ in this order,
are yielded by contracting to points all subgraphs of the diagrams involved, and so they are formed by the one-loop renormalization of the coupling $a_s$ (i.e.\ $\beta_0$) and composite vertex (i.e.\ $V_\text{X}^{(0)}$).
Collecting together all subgraph contractions related to the one-loop charge and vertex renormalization, one obtains
$\hat{\mathbf{P}}\left[\beta_0 V^{(0)}_\text{X}(x,y) d(y)\right]$.
The second kind of renormalization is generated by the contractions of the composite vertex at the two-loop level:
 $2\,\hat{\mathbf{P}} \Bigl[V^{(0)}_\text{X}(x,z) \otimes V^{(0)}_\text{X}(z,y) d(y) \Bigr]$. Notice that the former term is proportional to $\beta_0$, while the latter is not. The same pattern can be observed in all coefficients $\Pi^\text{X}_{2,j}$, which is an evident example of the $\beta$-expansion representation,  see e.g.\ \cite{Kataev:2016aib}:
\begin{align}
	\Pi^\text{X}_{2,j}(x,y) = \beta_0\, \Pi^\text{X}_{2[\beta],j}(x,y) + (\beta_0)^0\,\Pi^\text{X}_{2[0],j}(x,y).
\end{align}

In this paper, we calculate $\Pi^\text{X}_{2[\beta],j}$~--- the $\beta_0$ parts of the N$^2$LO correlators. These pieces might be expected to dominate in this order because of the relatively large value of $\beta_0$. In the vector case, harbingers of this dominance can be seen in the lowest Mellin moments of the correlator (see section \ref{sec:4}). It should also be noted that to obtain the $\beta_0$ parts of the three-loop correlators, it suffices to compute only two-loop--like topologies~--- the NLO diagrams modified with two-point one-loop quark insertions in gluon lines. Then the entire $\beta_0$ part can be restored unambiguously via a replacement $\displaystyle n_f \to -\frac32 \beta_0$.

We start with our results for the $\langle V\,V \rangle$ correlator that is important for applications and passes the most comprehensive independent test presented in section~\ref{sec:VVcorr-check} below. Then we turn to the $ \langle T\,T \rangle$ and $\langle S\,S \rangle$ correlators.


\subsection{\boldmath $\langle V\,V \rangle(x,y)$ correlator }
 \label{sec:VVcorr}

Explicit expressions for the $\beta_0$ piece $\Pi_{2[\beta]}^\text{V}$ of the vector-vector correlator at N$^2$LO are given by the following formulae:
\begin{subequations}
 \label{eq:PiV2}
 \begin{eqnarray}
 \Pi^\text{V}_{2[\beta]} &=&-\frac{ N_c}{2\pi^2} C_F \sum_{j=0}^{3} \tilde\Pi^\text{V}_{2,j}(x,y) L^j;\\
\tilde{\Pi}^\text{V}_{2,3}&=&\frac{1}{3}W^{(0)}_+;  \label{eq:PiV23}\\
\tilde{\Pi}^\text{V}_{2,2}&=&-\frac{5}{3}W^{(0)}_+
+\left(\hat{\mathbf{S}}\theta(y>x)\left[ (\bar{y}+x)\ln(y-x)-(1-x\bar{y}-y\bar{x})\ln(y\bar{x})\right] \right)_+ \nonumber \\ && + \left( \ln|y-x|~ W^{(0)}_b\right)_+ + \frac{1}{2}\delta(y-x)\hat{\mathbf{S}}\left[x\ln(x)\right]
; \label{eq:PiV22}
\end{eqnarray}
 \begin{eqnarray}
\tilde{\Pi}^\text{V}_{2,1}&=&\frac{13}{9} W^{(0)}_{+}+\frac{2}{3} \dot{W}^{(0)}_{+}+\frac32 \ddot{W}^{(0)}_{+} - 2\mathop{\hat{\mathbf{S}}} \theta(y>x)
 \left[ (x+\bar{y}) \ln \left(x \bar{y}\right) \left( \ln \left(x \bar{y}\right)+\frac{5}3\right) \right] \nonumber\\
&&{}+(1-x \bar y-y \bar x) \left[ -\ln ^2\left(x \bar{y}\right) +2\ln ^2\left(y \bar{x}\right)+\frac{10}{3}
\ln\left(x y \bar{x} \bar{y}\right) \right]
\nonumber\\
&&{}+ (1-\left| x-y\right| ) \ln\left| x-y\right| \left(\ln\left| x-y\right| -\frac{10}3\right)
- W^{(0)} \mathop{\hat{\mathbf{S}}}  \theta(y<x) \left( \frac{\pi^2}6 - \mathop{\mathrm{Li}_2}(z)\right)
\nonumber\\
&&{}+\frac12 \mathop{\hat{\mathbf{P}}} \left[ \left( \dot{V}^{(0)} \dot{d} +\frac{1}{2} V^{(0)} \ddot{d}+\frac{5}{3} V^{(0)} \dot{d} + V_a \dot{d} \right)_{+(x)} \right]
\nonumber\\
&&{}+\left( \ln^2\left| y-x\right| W_b-\frac{10}{3} \ln\left| y-x\right| W_b+\frac{5}{3} W_b-\frac{7}{3} \dot{W}_b-2\ddot{W}_b\right)_+
\nonumber\\
&&{}+ \delta(x-y)\left(x\bar{x}\ln(x) \ln\left(\bar{x}\right)-\frac{55}{6} d +\frac{9}{2} \dot{d} +\frac{3}2 \ddot{d} \right)\,, \label{eq:PiV21}
\end{eqnarray}
 \end{subequations}
where $\displaystyle \ddot{d}=\frac{d^2}{d\varepsilon^2}d(y;{\varepsilon})\Big\vert_{\varepsilon=0}$.

As it is expected, the leading-log term $\Pi^\text{V}_{2,3}(x,y)$ is proportional to a plus-distribution prescribed by the vector-current conservation, which means that  $\Pi^\text{V}_{2,3}(x,\underline{0})=0$. In addition, the leading-log term at this order is diagonalized by the same set of the Gegenbauer polynomials $\{C^{(3/2)}_n(y-\bar{y})\}$ as at order $a_s$.


\subsubsection{Mellin moments of \boldmath $\langle V\,V \rangle(x,y)$ as a check of the correlator}
 \label{sec:VVcorr-check}
Vetting our calculation of the correlator $\langle V\,V \rangle(x,y)$, we must compare its lowest Mellin moments with the results of refs.~\cite{Gracey:2009da, Chetyrkin:2010dx}. In doing so, we find the following linear combinations of the moments to agree with the previous calculations:\footnote{Note the different definition of the correlator in \cite{Gracey:2009da} --- it is a correlation function of two $V$-operators (not $V$ and $V^\dagger$ as in the present paper) which explains $-\bar{y}$ in eq.~\eqref{eq:gracey-moments}.}
\begin{align}\label{eq:gracey-moments}
	\int_0^1 dx \, x^n \int_0^1 dy\, (-\bar{y})^m \Pi^\text{V}(x,y;p^2)
\end{align}
for $(n,m)=(0,0)$, $(1,0)$, $(2,0)$, $(1,1)$, and $(2,2)$.

The relation $\Pi^\text{V}(\underline{0},\underline{0};p^2) = 2\Pi^\text{V}(\underline{1},\underline{0};p^2)$ confirmed by the explicit calculations of ref.~\cite{Gracey:2009da} is an immediate consequence of the symmetry $\Pi^\text{V}(x,\underline{0};p^2)=\Pi^\text{V}(\bar{x},\underline{0};p^2)$. As it is seen from eq.~\eqref{eq:currveccons}, the moments $\Pi^\text{V}(\underline{n},\underline{0})$ do not contain the highest possible power of $L$ allowed at a given order of perturbation theory. This is also confirmed in \cite{Gracey:2009da}.
Finally, it is important to note that the $\Pi^\text{V}_{2,2}(\underline{n},\underline{0})$ and $\zeta_3$ part of $\Pi^\text{V}_{2,1}(\underline{n},\underline{0})$ in the complete calculation in \cite{Gracey:2009da}
are proportional to $\beta_0$ for $n=0$, 1, 2.  This might hint at the dominance of the  $\beta_0$ contribution evaluated here, which is discussed in section~\ref{sec:rhoL} in connection with the meson DAs.


\subsection{\boldmath $\langle T\,T \rangle(x,y)$ correlator}
\begin{subequations}
\label{eq:PiT2}
The expansion  eq.~(\ref{eq:Pi_2}) for the tensor-tensor correlator reads
\begin{eqnarray}
 \Pi^\text{T}_{2[\beta]} &=& \frac{ N_c}{\pi^2} C_F \sum_{j=0}^{3} \tilde\Pi^\text{T}_{2,j}(x,y) L^j;\\
\tilde{\Pi}^\text{T}_{2,3}&=&\frac{1}{3}W^{(0)}_\text{T},  \label{eq:PiT23}\\
\tilde{\Pi}^\text{T}_{2,2}&=& \left(\ln\lvert x-y\rvert W_b -\frac{5}3 W_b + \mathop{\hat{\mathbf{S}}} \left[ \theta(\bar{z} > 0)  \ln ( \bar z ) \right] \right)_+
\nonumber\\
											&&{}+ \delta(x-y) \left( \mathop{\hat{\mathbf{S}}} \left[ x \ln(x) \right] + \frac{19}6 d -\frac12 \dot{d} \right), \label{eq:PiT22}\\
\tilde{\Pi}^\text{T}_{2,1}&=& \mathop{\hat{\mathbf{S}}} \biggl\{ \theta(x>y) \biggl[2 \ln \left(\bar{z}\right) \ln (x-y)-\ln^2\left(\bar{z}\right)-\frac{10}3 \ln \left(\bar{z}\right)
 - 2 y \bar{x} \biggr] \biggr\}
\nonumber\\
&&{}-W^{(0)}_\text{T} \mathop{\hat{\mathbf{S}}} \left[ \theta(x>y) \left( \frac{\pi^2}6 - \mathop{\mathrm{Li}_2}(z)\right) \right]+ \frac12 \mathop{\hat{\mathbf{P}}} \left[ \left( \dot{V}_b \dot{d} +\frac{1}{2} V_b \ddot{d} +\frac{5}{3} V_b \dot{d} \right)_{+(x)} \right]
\nonumber\\
&&{} + \left( \ln^2\left| y-x\right|  W_b-\frac{10}{3} \ln\left| y-x\right| W_b+\frac{28}{9} W_b-\frac{5}{3} \dot{W}_b-\frac12\ddot{W}_b\right)_+
\nonumber\\
&&{}+ \delta(x-y) \left( x\bar{x}\ln(x)\ln\left(\bar{x}\right)-\frac{397}{36} d + \frac{19}{6} \dot{d} -\frac12 \ddot{d} \right), \label{eq:PiT21}
\end{eqnarray}
 \end{subequations}
where all elements of the notation in the above formulae are defined in eqs.~\eqref{eq:VecV}, \eqref{eq:PiV1}, (\ref{eq:PiT1}), and (\ref{eq:PiV2}).


\paragraph{Check of the moments of \boldmath $\langle T\,T \rangle(x,y)$.}
Integrating eqs.~\eqref{eq:PiT1} and \eqref{eq:PiT2} over $x$ and $y$, we can get the twofold zeroth moment $\Pi^\text{T}(\underline{0},\underline{0})$ which was also obtained in ref.~\cite{Gracey:2009da} (see section 4.3 therein). The moment we calculated coincides with the one in \cite{Gracey:2009da}. In addition, the moment $\Pi^\text{T}(\underline{1},\underline{0})$ can be extracted from the results listed in section 4.8 of \cite{Gracey:2009da}. It is precisely one-half less than the two-fold zeroth moment, $\Pi^\text{T}(\underline{1},\underline{0}) = \frac12 \Pi^\text{T}(\underline{0},\underline{0})$, which is a corollary of mirror symmetry of the one-fold zeroth moment $\Pi^\text{T}(x,\underline{0})=\Pi^\text{T}(\bar{x},\underline{0})$.


\subsection{\boldmath $\langle S\,S \rangle(x,y)$ correlator}

The expansion for the scalar-scalar correlator reads
\begin{subequations}\label{eq:PiS2}
\begin{eqnarray}
\Pi^\text{S}_{2[\beta]} &=&-\frac{ N_c}{8\pi^2} C_F \sum_{j=0}^{3} \tilde\Pi^\text{S}_{2,j}(x,y) L^j;
\\
\tilde{\Pi}^\text{S}_{2,3}&=& \frac{1}{3} V^{(0)}_\text{S} =   \frac2{3}\left[\hat{\mathbf{S}} \frac{\theta (y>x)}{y-x} \right]_+ +\delta(y-x),  \label{eq:Pi23}\\
\tilde{\Pi}^\text{S}_{2,2}&=& \frac{1}{2} \Biggl[1 -\frac1{2} \left( \ln \lvert x-y \rvert +\frac53 \right) W_a^\text{S}
- \frac1{2} \mathop{\hat{\mathbf{P}}} \left\{ \left[ \left( \ln \lvert x-y \rvert +\frac83 \right) W_b^\text{S} \right]_{+(x)} \right\}
\nonumber\\
&&{}\hphantom{\frac{1}{12} \Biggl[ }+ \delta (x-y) \left( \frac{\dot{d}_\text{S}}{2}- \frac{11}3 \right) \Biggr], \label{eq:PiS22}
\end{eqnarray}
\begin{eqnarray}
\tilde{\Pi}^\text{S}_{2,1}&=& -\frac{1}{4} \ddot{W}_a^\text{S}-\frac{5}{6} \dot{W}_a^\text{S} +  \left(\frac1{2} \ln ^2(\left| x-y\right| )-\frac5{6} \ln (\left| x-y\right| )+\frac{14}{9}\right) W_a^\text{S}
\nonumber\\
&&{}+\mathop{\hat{\mathbf{P}}} \left\{ \frac{1}{2} \dot{\tilde{V}}^\text{S}_a \dot{d}_\text{S} + \frac{1}{12} \left(3 \ddot{d}_\text{S} +10 \dot{d}_\text{S}\right) \tilde{V}^\text{S}_a \right\}
+\mathop{\hat{\mathbf{P}}}  \biggl\{ \biggl[ \frac{1}{36} \Bigl(-9 \ddot{d}_\text{S}+30 \dot{d}_\text{S}+18 \ln ^2(\left| y-x\right| )
\nonumber\\
&&{}-96 \ln (\left| y-x\right| )+152 \Bigr) W_b^\text{S}
+\frac{1}{6} \left(3 \dot{d}_\text{S} -5\right)  \dot{W}_b^\text{S} -\frac{1}{4} \ddot{W}_b^\text{S} \biggr]_{+(x)}\biggr\}
\nonumber\\
&&{} - W^{(0)}_\text{S} \mathop{\hat{\mathbf{S}}} \left[ \theta(\bar{z}>0) \left( \frac{\pi^2}6 - \mathop{\mathrm{Li}_2}(z)\right) \right]
+2 -\frac{3}{2} \ln \left(x \bar{x} y \bar{y}\right) -2 \mathop{\hat{\mathbf{S}}} \left[ \theta (\bar{z}>0) \ln \left(\bar{z} \right) \right]
\nonumber\\
&&{} + \delta(x-y) \left( \frac{19}{6}  \ln \left(x \bar{x}\right) -\frac{547}{36} -\frac12 \mathop{\hat{\mathbf{S}}} \left[ \ln ^2\left(x\right) \right] + \mathop{\hat{\mathbf{S}}} \left[ x \ln (x) \right] \right). \label{eq:PiS21}
\end{eqnarray}
  \end{subequations}
Here, $d_\text{S}$, $\dot{d}_\text{S}$, and $\ddot{d}_\text{S}$ were defined by eqs.~\eqref{eq:dS} and we have also introduced the generalized ``scalar'' kernels $\tilde{V}^\text{S}$ in analogy with the definitions in eqs.~(\ref{eq:VecV}) for the vector case
\begin{align}
	\tilde{V}_a^\text{S}(x,y\mid\varepsilon)
	= 2 \mathop{\hat{\mathbf{S}}} \!\left[ \frac{\theta (y>x)}{y} \left(\frac{x}{y}\right)^{\varepsilon} \right]\,,
~\tilde{V}^\text{S}_b(x,y\mid\varepsilon) = 2\hat{\mathbf{S}}\left[\frac{\theta(y>x)}{(y-x)}
\left(\frac{x }{y}\right)^{1+\varepsilon}\right]_+,
\end{align}
and
\begin{align}
W_a^\text{S}(x,y\mid\varepsilon) &{}\!=\! 2 \mathop{\hat{\mathbf{P}}} \! \mathop{\hat{\mathbf{S}}}\!\!\left[\frac{\theta (y>x)}{y}\left( x \bar{y} \right)^{\varepsilon} \right],\,
W_b^\text{S}(x,y\mid\varepsilon)
\!=\! 2 \mathop{\hat{\mathbf{S}}}\!\!\left[\frac{\theta (y>x)}{y-x} \frac{x}{y} \left(x\bar{y}\right)^{\varepsilon} \right], \\
W_{(0)}^\text{S}(x,y\mid\varepsilon)
&{}=
W_a^\text{S}(x,y\mid\varepsilon) + W_b^\text{S}(x,y\mid\varepsilon),
\end{align}
In eqs.~\eqref{eq:PiS2} as throughout this paper, dots over functions without arguments designate the coefficients of the corresponding Taylor series in $\varepsilon$, e.g.\
\begin{align}
W_{I}^\text{S}(x,y\mid\varepsilon)
=
W_I^\text{S} + \varepsilon \dot{W}_I^\text{S} + \frac1{2!} \varepsilon^2 \ddot{W}_I^\text{S} + \dots,
\qquad
I = (0),\,a,\,b.
\end{align}
Note here that $\tilde{\Pi}^\text{S}_{2,3}(x,y)$ is diagonalized by the set $\{ C^{(1/2)}_n(\bar{y}-y ) \}$
due to the expected property $\tilde{\Pi}^\text{S}_{2,3}(x,y) \sim V^{(0)}_\text{S}(x,y)$.


\textbf{Check of the moments of $\langle S\,S \rangle(x,y)$.}
If we evaluate the double zeroth moment $\Pi^\text{S}(\underline{0},\underline{0})$ integrating the correlator $\Pi^\text{S}(x,y)$ over $x$ and $y$, the result coincides with the calculation of ref.~\cite{Gracey:2009da} (see section 4.1 therein).


\section{Radiative content of  meson DAs within QCD sum rules}
 \label{sec:4}

In this section, we apply our results for the correlators to the description of exclusive hard hadron processes in terms of DAs. Technically, these DAs are linked to the moments $\Pi^\text{X}(x,\underline{0};P^2)$ and $\Pi^\text{X}(\underline{a},\underline{0};P^2)$, see the definitions in (\ref{def: moments-ab}).
These moments are obtained from the correlators of two composite vertices, $\Pi^\text{X}(x,y;P^2)$, presented in sections~\ref{sec:2} and \ref{sec:3}. The expressions for the moments were given in eqs.~(\ref{eq:NLOcorr}) and \eqref{eq:TTNLO} to two-loop order.
Here, we write down the final results up to order $\beta_0 a_s^2$ and focus on the perturbative content of the DAs to only estimate its effect,
while a full-fledged analysis of the DA properties in QCD SR will be given elsewhere.

Let us recall some elements of the Borel SR approach that is used to determine meson DAs.
This kind of SR is based on the dispersion relation  for the one-fold correlator
$\Pi^\text{X}(x,P^2) \equiv \Pi^\text{X}(x,\underline{0};P^2)$:
\begin{eqnarray}
\label{eq:disp_rel}
\Pi^\text{X}(x,P^2)= \frac1{\pi} \int^\infty_\text{thresh} \frac{\text{Im} \left[\Pi^\text{X}(x,s)\right]}{s+P^2}~ds~
+ \text{``subtractions'', }
\end{eqnarray}
where $\Pi^\text{X}$ is constructed with a current $J_\text{X}$ that has a nonvanishing projection on a meson state $M$ described by the corresponding DA, see the discussion and definitions in appendix~\ref{App:DA}.
The subtractions in the r.h.s. of the relation above can be polynomials in $P^2$.
To reinforce the contribution of the lowest-state meson in the r.h.s.\ and to improve the convergence in the l.h.s.,
one usually applies the Borel transform $\hat{\mathbf{B}}_{(M^2)}$,
\begin{subequations}
\begin{eqnarray}
\label{eq:borel-1}
\hat{\mathbf{B}}_{(M^2)}\left[ \Pi^\text{X}(x,P^2)\right] &=& \lim_{P^2 =n M^2,~ n \to \infty}
\frac{(-P^2)^n}{\Gamma(n)} \frac{d^n}{d (P^2)^n}\Pi^\text{X}(x,P^2),
\end{eqnarray}
 to both sides of \eqref{eq:disp_rel}, which leads to
\begin{eqnarray}
\hat{\mathbf{B}}_{(M^2)} \left[ \Pi^\text{X}(x,P^2) \right] = \frac1{\pi} \int^\infty_\text{threshold}\text{Im} \left[\Pi^\text{X}(x,s)\right] \exp\left(-\frac{s}{M^2}\right)~\frac{ds}{M^2} \,. \label{eq:borel-2}
\end{eqnarray}
\end{subequations}
The Borel transform ``kills'' all polynomials in $P^2$ in the r.h.s.\ saving only logarithmic terms $L^n$, $n \geqslant 1$ in the l.h.s.\ of (\ref{eq:disp_rel}).
Under this transform, any powers $L^n$, $n \in \mathbb{N}$ turn into a polynomial in
$L_\text{B} = \ln\left(\frac{M^2}{\mu^2}e^{-\gamma_\text{E}}\right)$, see eq.~(\ref{eq:E-lnp}) for the general case.
To transform the correlators at N$^2$LO, we need the following special cases:
\begin{align}
	\hat{\mathbf{B}}_{(M^2)}L^0=0,
	\quad
	\hat{\mathbf{B}}_{(M^2)}L=-1,
	\quad
	\hat{\mathbf{B}}_{(M^2)}L^2=-2 L_\text{B},
	\quad
	\hat{\mathbf{B}}_{(M^2)}L^3=-3\left(L_\text{B}^2-\frac{\pi^2}{6} \right).
\end{align}

Finally, it is instructive to note a useful and general property of the moments $\Pi^\text{X}(\underline{N},\underline{0};\linebreak[1] P^2)$.
All these moments (with $N$ being a natural number) correspond to local vertices.
They do not contain terms proportional to $\pi^2$ in agreement with Kotikov's and Baikov's conclusions
\cite{Kotikov:2019bqo,Baikov:2019zmy}.
At the same time, the inverse moment $\Pi^\text{X}(\underline{-1},\underline{0};P^2)$ contains the
$\pi^2$-term because the moment does not correspond to a local operator.

\subsection{\boldmath Radiative content of twist-2 DAs for $\pi$ and $\rho_L$ mesons}
\label{sec:rhoL}
\begin{figure}[ht]
\centering
\includegraphics[width=0.49\textwidth]{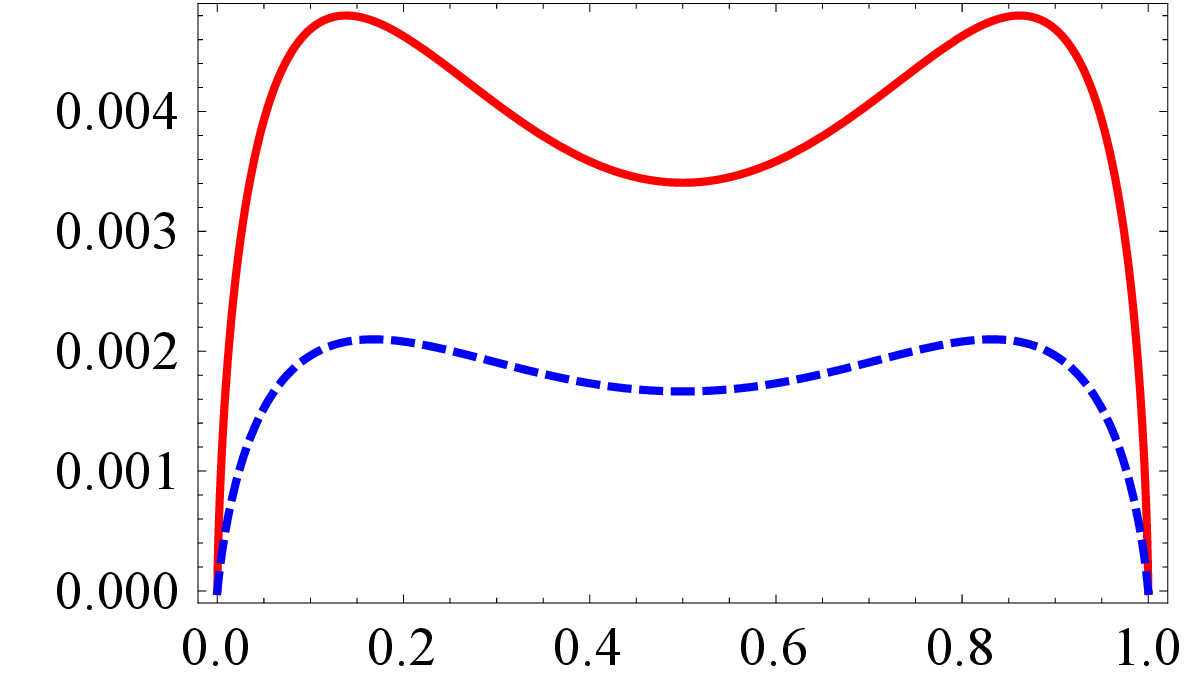}
\includegraphics[width=0.49\textwidth]{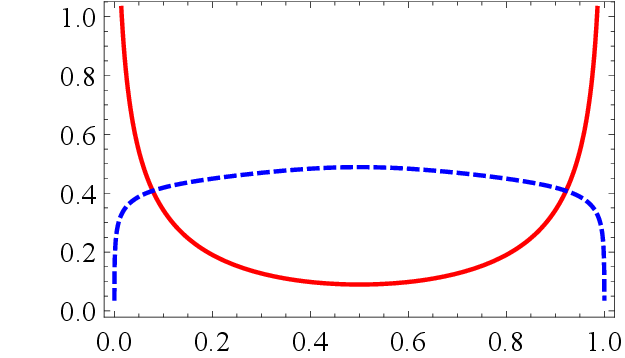}
\caption{\label{fig2}  Left panel: NLO \textcolor{red}{\textbf{(---)}} and $\beta_0$N$^2$LO \textcolor{blue}{$\bm{(- - - )}$} contributions to DAs for pseudoscalar or longitudinally polarized vector mesons, $N_c/(2\pi^2) \cdot a_s C_F \Delta \tilde\varphi^{(1)}_{M_{\parallel(A)}}$ and $N_c/(2\pi^2) \cdot a_s^2 \beta_0 C_F \Delta \tilde\varphi^{(2[\beta])}_{M_{\parallel(A)}}$ respectively, see eqs.~(\ref{eq:NNLOcorr}). Right panel: the ratios NLO/LO $=a_s C_F \Delta \tilde\varphi^{(1)}_{M_{\parallel(A)}}/\Delta \tilde\varphi^{(0)}_{M_{\parallel(A)}}$ \textcolor{red}{\textbf{(---)}} and $\beta_0$N$^2$LO/NLO
$= a_s\beta_0\Delta \tilde\varphi^{(2[\beta])}_{M_{\parallel(A)}}/\Delta \tilde\varphi^{(1)}_{M_{\parallel(A)}}$
 \textcolor{blue}{$\bm{(- - -)}$}. All curves are for the case of $L_\text{B}=0$, $\alpha_s (\mu^2 = 1 \text{ GeV}^2) \approx 0.494$. }
\end{figure}

Here, we start with $\Pi^\text{V}$ correlator that determines the perturbative part of $\pi$ and $\rho_L$ meson DAs.
Integrating eq.~\eqref{eq:PiV2} over $y$, taking its Borel transform, and combining the result with eq.~(\ref{eq:NLOcorr}), we arrive at
\begin{subequations}\label{eq:NNLOcorr}
\begin{align}
\Delta \varphi^{(0+1+2)}_{M_{\parallel(A)}}(x;M^2)& =
     \hat{\mathbf{B}}_{(M^2)}\Pi^\text{V}_{\text{0+1+2}}(x;P^2)
 \nonumber \\
&{}= \frac{N_c}{2\pi^2} \biggl\{ \Delta \tilde\varphi^{(0)}_{M_{\parallel(A)}}
			+ a_s C_F \Delta \tilde\varphi^{(1)}_{M_{\parallel(A)}}
			+ a_s^2 \beta_0 C_F \Delta \tilde\varphi^{(2[\beta])}_{M_{\parallel(A)}} + a_s^2 C_F \Delta \tilde\varphi^{(2[0])}_{M_{\parallel(A)}}\label{eq:totalNNLOcorr}
			\biggr\},
\end{align}
\begin{align}
	\Delta \varphi^{(0)}_{M_{\parallel(A)}} = {}& x \bar{x},
\\
	\Delta \tilde\varphi^{(1)}_{M_{\parallel(A)}} = {}& x \bar{x} \left[ 5-\frac{\pi^2}{3}+\ln^2\left(\frac{\bar{x}}{x}\right) \right],	 \label{eq:x0-NLO}
\\
	\Delta \tilde\varphi^{(2[\beta])}_{M_{\parallel(A)}} = {}& \hat{\mathbf{S}} \biggl[ - x \bar{x} \biggl( 5 \Li{3}(x) - \ln x \, \Li{2}(x) + \frac12 \ln x \, \ln^2{\bar{x}} -\frac{5}{12} \ln^2\frac{\bar{x}}{x} -\frac{1}{6} \ln^3{x} -\frac{\pi^2}{3} \ln{x}
	 \nonumber\\ \label{eq:cor2}
{}& \hphantom{\hat{\mathbf{S}} \biggl[ - x \bar{x} \biggl( }
+ \frac{5 \pi^2}{36}   -\frac{7}{12}\biggr) - x \left( \Li{2}(x)-\frac{\pi^2}6 - \frac34 \ln^{2}{x} + \frac{31}{12}\ln{x}  -L_\text{B}\ln{x} \right)\biggr],
\end{align}
 \end{subequations}
 which has been already presented in the proceedings \cite{Mikhailov:2020izb}, while the last term $\Delta \tilde\varphi^{(2[0])}_{M_{\parallel(A)}}$ in (\ref{eq:totalNNLOcorr}) is still unknown.
 The perturbative part $\Delta \varphi_{M}$ is common for twist-2 DAs of both $\pi$ and $\rho_{\parallel}$ mesons.
As we can see in figure~\ref{fig2}, the impact of the contribution \eqref{eq:cor2} of order $a_s^2\beta_0$ looks especially significant for intermediate values of $x$ and less important in the vicinity of endpoints.

Phenomenologically important characteristics of $\Delta \varphi_{M}$ are its norm and normalized moments defined as
\begin{gather}
	\langle f(x) \rangle_{M}
	\equiv
	\left. \left[ \int_0^1 dx \, f(x) \Delta \varphi_{M}(x) \right] \middle/ \left[ \int_0^1 dx \, \Delta \varphi_{M}(x) \right] \right.,
\\ \label{eq:Vx^0}
	\mathscr{N} = \int_0^1 dx \, \Delta \varphi_{M}(x)
=
	\frac{N_c}{12\pi^2}
	\Biggl\{
		1
		+ a_s C_F 3
		+ a_s^2 C_F \left[ C_A -\frac32 C_F + \beta_0 3 \left( \frac{11}{2} - 4 \zeta_3 - L_\text{B} \right) \right]
	\Biggr\}.
\end{gather}
In particular, we are interested in the inverse and second $\xi$ moment, $\xi = 2 x - 1$:
\begin{align}\notag
\langle x^{-1} \rangle_{M_{\parallel}} ={}& \frac1{\mathscr{N}_0} \frac{ N_c}{4\pi^2}\left[ 1 + a_s C_F 5+  a_s^2 \beta_0 C_F 3 \left( \frac{7}{18}-\frac{5}{3} \zeta_3+ \frac{31}{108} \pi^2 -\frac{\pi ^2}{9}  L_\text{B} \right)\right]
\\
{} \approx {}& \frac1{\mathscr{N}_0} \frac{ N_c}{4\pi^2}\left[ 1 + a_s C_F 5+  a_s^2 \beta_0 C_F 3 \left( 1.2184 - 1.0966 L_\text{B} \right)\right], \label{eq:x^-1}
\end{align}
\begin{equation}
\langle \xi^2 \rangle_{M_{\parallel}} =
\frac1{\mathscr{N}} \frac{N_c}{60 \pi ^2}  \Biggl\{ 1 + a_s C_F  5
+ a_s^2 C_F \left[ \frac{1}{72} C_A + \frac{353}{72} C_F + \beta_0 \left( \frac{1327}{72}-12 \zeta_3 -
\frac{10}{3}  L_\text{B}\right) \right] \Biggr\},
\label{eq:Vxi2}
\end{equation}
where $\mathscr{N}_0$ is the norm \eqref{eq:Vx^0} with the $(\beta_0)^0$ piece being omitted in order $a_s^2$ since only the $\beta_0$ part of the inverse moment has been calculated up to date. The norm \eqref{eq:Vx^0} is essentially the Adler $D$-function (up to a factor).

In eqs.~\eqref{eq:Vx^0} and \eqref{eq:Vxi2}, we have extracted the $(\beta_0)^0$ pieces from the correlators in ref.~\cite{Gracey:2009da}.
It is worth stressing again that all other terms of the norm and $\xi^2$ moment calculated by us coincide with those that can be extracted from ref.~\cite{Gracey:2009da}.

The $\Delta \varphi^{(2[\beta])}_{M_{\parallel(A)}}$ in (\ref{eq:cor2}) makes a minor
contribution to the inverse moment with respect to lower orders~--- compare the third term and the second one in eq.~(\ref{eq:x^-1}), their ratio is $0.085$ for $L_\text{B}=0$ and $\alpha_s (\mu^2 = 1 \text{ GeV}^2) \approx 0.494$. This $\beta_0$ part, however, is known to dominate the norm \eqref{eq:Vx^0} numerically in order $a_s^2$.\footnote{This observation was a reason to invent the BLM optimization \cite{Brodsky:1982gc}.}
It is instructive to verify numerical validity of large-$\beta_0$  approximation for the
$\langle \xi^2 \rangle_{M_{\parallel}}$ moment comparing it with the exact expression that can be obtained from the complete calculations in \cite{Gracey:2009da},
\begin{align}
\langle \xi^2 \rangle_{M_{\parallel}} =
\frac1{\mathscr{N}} \frac{N_c}{60 \pi ^2}  \left\{ 1 + a_s C_F  5 + a_s^2 C_F \left[ 6.5787 + \beta_0
\left(  4.0059 - \frac{10}{3}  L_\text{B}\right) \right] \right\} . \label{eq:Vxi2tot}
\end{align}
It is easy to see that the $\beta_0$ part is dominant in this moment also (at $L_\text{B}\approx 0$).
In addition, we can estimate the perturbative QCD contribution to the Gegenbauer moment $a_2$,
although one should recognize that a significant contribution to $a_2$ could come from nonperturbative vacuum-condensate
interactions that can vary depending on quantum numbers of mesons.
The perturbative contribution $a^{\parallel r}_2$ ($r$ stands for ``radiative'') is proportional exactly to $a_s(\mu^2)$:
\begin{eqnarray}
 a^{\parallel r}_2 &=&\frac7{12}\left(5\langle \xi^2 \rangle_{M_{\parallel}}-1  \right) \nonumber  \\
 &=&a_s C_F \frac{7}{6} \frac{ 1 + a_s \left[ \underline{2.79} + \beta_0 \left( 0.97- 0.1(6)L_\text{B} \right) \right]}{1 + a_s C_F 3 + a_s^2 C_F
  \left[ \underline{1} + \beta_0 3 \left(0.69 - L_\text{B} \right) \right]}\Big|_{L_\text{B}=0} \approx 0.069~(\underline{0.074}), \label{eq:a2par-NNLO}
\end{eqnarray}
where we have set $\mu^2=1$~GeV$^2$, $\alpha_s(\mu^2)=0.494$, and $L_\text{B}=0$ in the r.h.s.; the first value is obtained with the (underscored) non-$\beta_0$ parts neglected at N$^2$LO, while the second (underscored) one is exact with accounting for all terms.
The condition $L_\text{B}=0$ is compatible with the ``stability window'' of the corresponding QCD SR for the Borel parameter $M^2$.

It is useful to compare the estimate (\ref{eq:a2par-NNLO}) with
\begin{enumerate}
\item
QCD SR results: $a_{2}^{\parallel\rho}=0.047(58) < a^{\parallel r}_2=0.069\;  (\underline{0.074})< a_{2}^{\pi}=0.187(60)$ \cite{Pimikov:2013usa, Stefanis:2015qha};
\item
lattice results: $a_2^{\parallel\rho}=0.184(18)(33)$, $a_{2}^{\pi}=0.140(24) > a^{\parallel r}_2=0.069\; (\underline{0.074})$
at $\mu^2=1$~GeV$^2$ (which are evolved from the values at $\mu^2=4$~GeV$^2$ in \cite{Bali:2019dqc,Braun:2016wnx}).
\end{enumerate}

As we can see, the radiative contribution $a^{\parallel r}_2$ is of the same order of magnitude as the complete $a_2$,
so that the contribution $a^{\parallel r}_2$ is comparable numerically with the nonperturbative one and, therefore, is important to take it into account.


\subsection{Radiative content of \boldmath $\rho_T$-meson twist-2 DAs}

From the tensor correlator \eqref{eq:PiT2} we get a next-order correction to the NLO amplitude (\ref{eq:TTNLO}):
\begin{subequations}\label{eq:TNNLOcorr}
\begin{align}
\Delta \varphi^{(0+1+2)}_{M_{\perp}}(x;M^2) & =
    \hat{\mathbf{B}}_{(M^2)}\Pi^\text{T}_{0+1+2}(x;P^2)
 \nonumber \\
&{}= -\frac{N_c}{\pi^2} \biggl\{ \Delta \tilde\varphi^{(0)}_{M_{\perp}}
			+ a_s C_F \Delta \tilde\varphi^{(1)}_{M_{\perp}}
			+ a_s^2 \beta_0 C_F \Delta \tilde\varphi^{(2[\beta])}_{M_{\perp}} + a_s^2 C_F \Delta \tilde\varphi^{(2[0])}_{M_{\perp}}\label{eq:totalNNLOcorrT}
			\biggr\},
\end{align}
\begin{align}
	\Delta \varphi^{(0)}_{M_{\perp}} = {}& x \bar{x},
\\
	\Delta \tilde\varphi^{(1)}_{M_{\perp}} = {}& x \bar{x} \left[6-\frac{\pi^2}{3}+\ln^2\left(\frac{\bar{x}}{x}\right)+\ln(x\bar{x})+2L_\text{B}\right],	 \label{eq:TNLO}
\\
	\Delta \tilde\varphi^{(2[\beta])}_{M_{\perp}} = {}& \frac{1}{6} \hat{\mathbf{S}} \biggl\{ 3 x \bar{x} \left( \frac{\pi^2}{6} -L_\text{B}^2 \right) + x \left[ 6 \left( 2 - \bar{x} \right) \ln \left(x\right) +19 \bar{x} \right]  L_\text{B}  - x \bigl[ 12 \text{Li}_2(x) -2 \pi^2
	\notag\\
	&{}  +16 \ln(x) - 9 \ln^2(x) \bigr]
+x \bar{x}\biggl[-30 \text{Li}_3(x) +6 \text{Li}_2(x) \ln(x) + \ln^3(x) -5 \ln(x) \ln(\bar{x})
 \nonumber \\
& {} + \ln^2(x) \left[ 2 - 3 \ln(\bar{x}) \right]+ \left( 2 \pi^2 + 19 \right) \ln(x) -\frac{5 \pi^2}{6}-\frac{193}{12} \biggr] \biggr\},
\label{eq:TNNLOb0}
\end{align}
\end{subequations}

\begin{figure}[ht]
\centering
\includegraphics[width=0.49\textwidth]{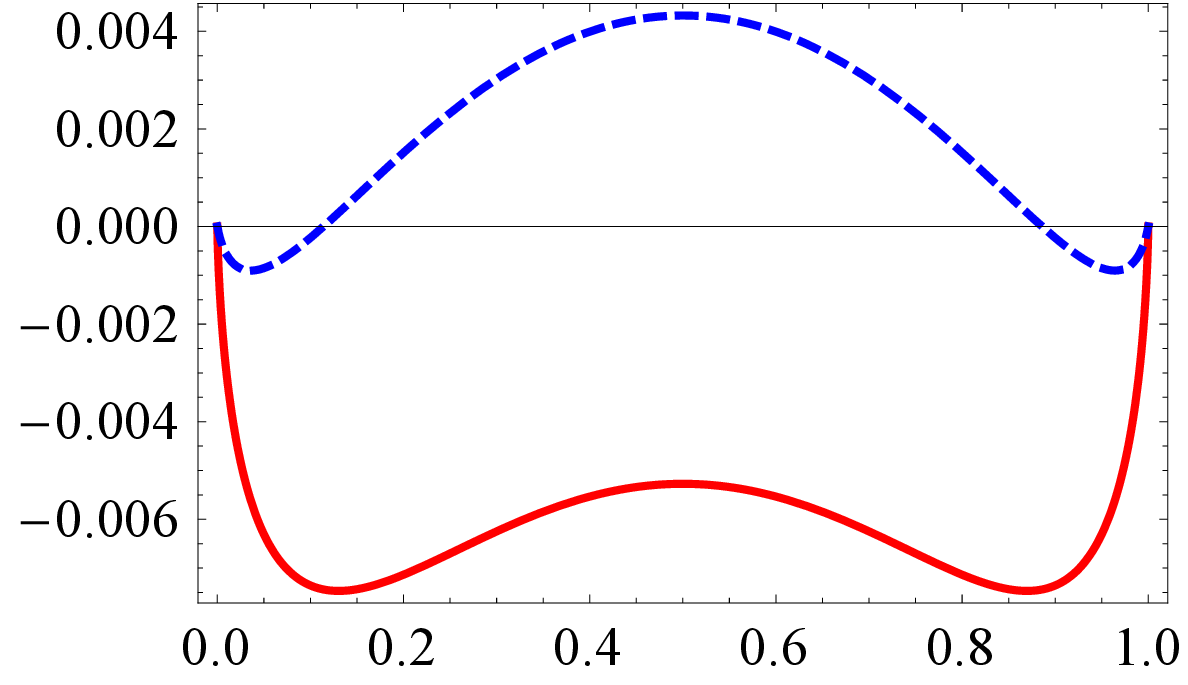}
\includegraphics[width=0.49\textwidth]{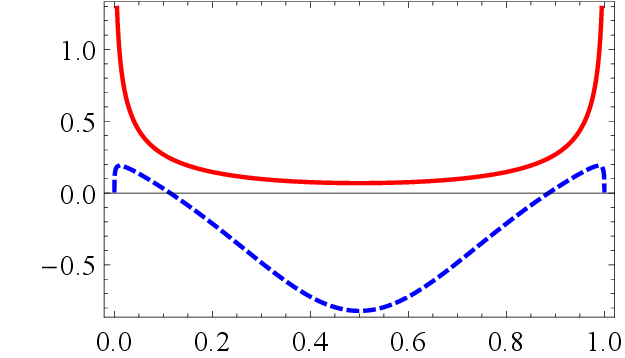}
\caption{\label{fig3} Left panel: NLO \textcolor{red}{\textbf{(---)}} and $\beta_0$N$^2$LO \textcolor{blue}{$\bm{(- - - )}$} contributions to DAs for transversally polarized vector mesons, $-N_c/\pi^2 \cdot a_s C_F \Delta \tilde\varphi^{(1)}_{M_\perp}$ and $-N_c/\pi^2 \cdot a_s^2 \beta_0 C_F \Delta \tilde\varphi^{(2[\beta])}_{M_{\perp}}$ respectively, see eqs.~\eqref{eq:TNNLOcorr}. Right panel: the ratios NLO/LO $=a_s C_F \Delta \tilde\varphi^{(1)}_{M_{\perp}}/\Delta \tilde\varphi^{(0)}_{M_{\perp}}$ \textcolor{red}{\textbf{(---)}} and $\beta_0$N$^2$LO/NLO
$= a_s\beta_0\Delta \tilde\varphi^{(2[\beta])}_{M_{\perp}}/\Delta \tilde\varphi^{(1)}_{M_{\perp}}$
 \textcolor{blue}{$\bm{(- - -)}$}. All curves are for the case of $L_\text{B}=0$, $\alpha_s (\mu^2 = 1 \text{ GeV}^2) \approx 0.494$.}
\end{figure}

In comparison with the LO and NLO terms, the $\beta_0$ part of the N$^2$LO contribution is mainly of the opposite sign
and comparable in magnitude with NLO  in the middle region of $x$, see figure~\ref{fig3}.

The norm, the inverse and $\xi^2$ moments of $\Delta \varphi^{(0+1+2)}_{M_{\perp}}$ read
 \begin{subequations}
 \begin{align}
&\mathscr{N} = \int_0^1 dx \, \Delta \varphi_{M_{\perp}}(x) = -\frac{N_c}{6\pi^2} \Biggl\{ 1 +
  a_s C_F \left( \frac73 + 2 L_\text{B} \right) \nonumber \\
  &{}+ a_s^2 C_F \biggl[ C_A\left(14\zeta_3 - \frac{407}{18}+\frac{38}{3}L_\text{B}\right)+C_F\left(\frac{1075}{36}-\frac{\pi^2}{3}-4\zeta_3+2L_\text{B}^2
  -\frac{43}3 L_\text{B}\right) \label{eq:TTN2LOx^0compl1} \\
  & \hphantom{{}+a_s^2 C_F \biggl[ } {} + \beta_0  \left( \frac{\pi^2}{6} -12 \zeta_3 + \frac{383}{36} + 2 L_\text{B} - L_\text{B}^2 \right) \biggr]\Biggr\}, \label{eq:TTN2LOx^0compl2}
\end{align}
\begin{align}
\langle \xi^2 \rangle_{M_{\perp}} &{} =
-\frac1{\mathscr{N}_0} \frac{N_c}{30\pi^2} \Biggl[ 1 + a_s C_F \left(2 L_\text{B}+\frac{59}{15}\right)
  \notag \\
& \hphantom{{} = \frac1{\mathscr{N}_0} \frac{N_c}{60\pi^2} \Biggl[}
{} + a_s^2 \beta _0 C_F
   \left(\frac{26}{15} L_\text{B}-L_\text{B}^2-12 \zeta_3 +\frac{\pi^2}{6}+\frac{1207}{100}\right) \Biggr],
\end{align}
\begin{align}
\langle x^{-1} \rangle_{M_{\perp}} = -\frac1{\mathscr{N}_0}\frac{N_c}{2\pi^2} \Biggl[ & 1 + a_s C_F \left(4 + 2 L_\text{B} \right)
	 \nonumber \\
&{}+ a_s^2 \beta_0 C_F \left( 2 \zeta_3 + \frac{19 \pi ^2}{18} - \frac{493}{36} + \frac{25-2\pi^2}{3} L_\text{B} - L_\text{B}^2 \right) \Biggr],\label{eq:TTN2LOx^-1}
\end{align}
  \end{subequations}
where $\mathscr{N}_0$ is the norm \eqref{eq:TTN2LOx^0compl2} with the $(\beta_0)^0$ piece \eqref{eq:TTN2LOx^0compl1} omitted,
the latter one can be obtained using the results of ref.~\cite{Gracey:2009da}.
The $\beta_0$ part of the norm (\ref{eq:TTN2LOx^0compl2}) is larger in magnitude and has the opposite sign in comparison
with the sum of non-$\beta_0$ terms in (\ref{eq:TTN2LOx^0compl1}) at $L_\text{B}=0$.
So the $\beta_0$ approximation works satisfactorily here,
although it is not as reliable in this case as in the vector one.

A significant convexity in the $x$-behavior of the $\beta_0$ part occurs in the middle values of $x$.
The negative NNLO contribution to $\langle x^{-1} \rangle_{M_{\perp}}$ in eq.~(\ref{eq:TTN2LOx^-1}) is not strong in comparison with the NLO one.
The radiative contribution $a^{\perp r}_2$ to $a_2^{\perp\rho}$ can be estimated in analogy to eq.~(\ref{eq:a2par-NNLO}),
 \begin{align}
  \label{eq:a2perp}
 a^{\perp r}_2  &{} \approx
a_s C_F \frac{14 }{15}\frac{ 1 + \beta _0 a_s \left(0.89-0.1(6)L_\text{B}\right)}{1 + a_s C_F \left( 2.(3) + 2 L_\text{B} \right)
	- a_s^2 \beta_0 C_F  \left(2.14 - 2 L_\text{B} + L_\text{B}^2 \right) }\Big|_{L_\text{B}=0}
\approx
0.05949\,.
\end{align}
The estimate in the r.h.s. of (\ref{eq:a2perp}) is obtained for $L_\text{B}=0$, $\mu^2=1$~GeV$^2$, $\alpha_s(\mu^2)=0.494$.
At these conditions, $a^{\perp r}_2=0.059<$ $a_2^{\perp\rho}=0.130$ from the lattice results \cite{Braun:2016wnx}
(originally, $a_2^{\perp\rho}=0.101(22)$ at $\mu^2=4$~GeV$^2$).
 Again, the radiative contribution $a^{\perp r}_2$ is large and as important as for the vector (axial) case.


\subsection{Radiative content of \boldmath $\pi$-meson twist-3 DA}

Having further applications to QCD SRs in mind, we have obtained the one-fold scalar correlator $\Pi^\text{S}_{\text{0+1+2}}(x,\underline{0};P^2)$ by integrating eqs.~(\ref{eq:PiS0}), \eqref{eq:PiS1}, and \eqref{eq:PiS2}.
This correlator is given explicitly in appendix~\ref{App:C}, eq.~(\ref{eq:PiS3}).
Below, we present the Borel transform of this correlator order by order:
\begin{subequations}
\begin{align}
\Delta\varphi^{p(0+1+2)}_{3;\pi}(x;M^2)& =
     \hat{\mathbf{B}}_{(M^2)} \left[ -P^2 \Pi^\text{S}_{\text{0+1+2}}(x,\underline{0};P^2) \right]
 \nonumber \\
&{}= \frac{N_c}{8\pi^2} M^2 \biggl\{ 1
			+ a_s C_F \Delta \tilde\varphi^{p}_{\pi,1}
			+ a_s^2 \cdot \left( \beta_0 C_F \Delta \tilde\varphi^{p}_{\pi,2[\beta]} + C_F\Delta \tilde \varphi^{p}_{\pi,2[0]} \right)
			\biggr\}
\end{align}
\begin{align}
	\Delta \tilde\varphi^{p}_{\pi,1} = {}& 5 - 3 \ln(\bar{x}x) - 6 L_\text{B},	
\\
	\Delta \tilde\varphi^{p}_{\pi,2[\beta]} = {}& 3 L_\text{B}^2 + \left[ \ln (x \bar{x} ) - 14 \right] L_\text{B}
	 - \mathop{\hat{\mathbf{S}}} \Bigl[ 5 \Li{3}(x) - \Li{2}(x) \ln(x) + x \ln (x) \Bigr] +\frac{1}{6} \ln^3(x\bar{x})
\notag\\
	&{} + \frac43 \ln^2\left( \frac{x}{\bar{x}} \right) - \ln (x) \ln(\bar{x}) \left[ \ln(x \bar{x}) + 1 \right]
	+ \left(\frac{\pi ^2}{3}-\frac{25}{6}\right) \ln(x \bar{x} ) -\frac{10}{9} \pi ^2+\frac{239}{12}.
\end{align}
 \end{subequations}

\section{Conclusion}
\label{sec: concl}
Here, we have calculated the massless correlators $\Pi^\text{V,T,S}(x,y; p^2)$ of two vector, tensor, and scalar composite vertices with the Bjorken fractions $x$ and $y$ at orders $\alpha_s$ and $\alpha_s^2 \beta_0$ of QCD. These correlators are universal objects appearing as a result of the collinear factorization procedure in hard processes.
We have discussed in detail the  structure of  the correlators and its elements and their relation to generalized
ERBL evolution kernels.
Moreover, we have verified our results by comparing them with the known particular cases for Mellin moments.
These results are used to estimate the impact of the radiative corrections  following from $\int_0^1\Pi^\text{X}(x,y;p^2)dy$ on distribution amplitudes of different  light mesons within QCD sum-rule approach.
For all cases, these radiative corrections  are significant and should be taken into account in DA calculations.

\acknowledgments
We would like to thank A.~Pikelner for bringing ref.~\cite{Chetyrkin:2010dx} to our notice and N.~Stefanis for clarifying discussions. The work of NV was supported by a grant of the Russian Science Foundation (Project No-18-12-00213).

\appendix

\section{Feynman rules for composite vertices}
\label{app:A}

The Feynman rules for composite vertices with $K$ gluon partons can be written as follows (all parton momenta are incoming with respect to the vertex):
\begin{gather*}
\raisebox{-58pt}{\includegraphics[width=0.45\textwidth]{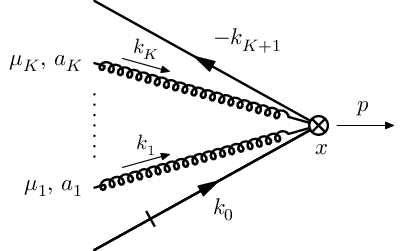}}\quad {}
=
g^K \Gamma^{\bar\nu}_\text{X} \,  \left( \prod_{i=1}^K \tilde{n}_{\mu_i} \right) \sigma^{a_1 \dots a_K}_K(x;x_0,x_1,\dots ,x_K).
\end{gather*}
Here, $\Gamma^{\bar\nu}_\text{X}$, $\text{X} = \text{S}$, P, V, A, T is the tensor-matrix structure defined by eq.~\eqref{eq:currents-def2}; $a_i$, $i=0,\dots K+1$ are SU(3) gluon indices; $\tilde{n}_\mu$ is a light-like vector normalized so that $\tilde{n} p = 1$; $x_i = \tilde{n} k_i$, $i=0,\dots K+1$ are longitudinal parton momentum fractions, $\sum x_i = 1$; and $\sigma^{a_1 \dots a_K}_K(x;x_0,x_1,\dots ,x_K)$ is a linear combination of the Dirac delta functions. Up to order $a_s^2 \beta_0$, we need composite vertices with no more than one gluon leg ($K=0$, 1). The corresponding expressions are given explicitly below, while higher-order vertices ($K \geqslant 2$) can be obtained by recursion:
\begin{align}
\sigma_0(x;x_0) &{}= \delta\left(x - x_0 \right),
\\
\sigma_1^{a_1}(x;x_0,x_1) &{}= t_{a_1} \frac{1}{x_1} \left[ \delta\left( x - x_0 - x_1 \right) - \delta\left( x - x_0 \right) \right],
\\ \notag
\sigma^{a_1 \dots a_K}_K(x;x_0,x_1,\dots ,x_K) &{}= \hat{\mathbf{S}}_{1,\dots,K} \biggl\{ t_{a_K} \cdots t_{a_2}
\sum_{i=0}^{K-1} (-)^i \sigma_1^{a_1}(x;x_{0\dots K-i-1},x_{K-i\dots K})
\\ &{} \hphantom{{}=\hat{\mathbf{S}}_{1,\dots,K} \biggl\{ t_{a_K} \cdots t_{a_2}}
\times \biggl(  \prod_{j=1}^{K-i-1} \frac{1}{x_{j \dots K-i-1}} \biggr) \biggl( \prod_{j=K-i}^{K-1} \frac{1}{x_{K-i \dots j}} \biggr)
 \biggr\},
\end{align}
where $t_a = \lambda_a / 2$, $x_{i_1 \dots i_k} = x_{i_1} + \dots + x_{i_k}$, and $\hat{\mathbf{S}}_{i_1,\dots i_n}$ designates total symmetrization of the indices $i_1,\dots i_n$ of the Gell-Mann matrices and the fractions $x_i$, e.g.\ $\hat{\mathbf{S}}_{1,2} f(1,2) = f(1,2) + f(2,1)$.

To simplify practical calculations involving the vertices given above, it makes sense to get rid of the denominators linear in parton momenta by introducing auxiliary integrations with the help of the Dirac deltas \cite{Mikhailov:1985cm}, e.g.
\begin{gather*}
\sigma_1^{a_1}(x;x_0,x_1)
=
t_{a_1} \int_0^1 \int_0^1 \mathrm{d}y_1 \, \mathrm{d}y_2\, \Theta(y_1+y_2<1) \rho_1(x;y_1,y_2) \delta\left(y_1 - x_0 \right)  \delta\left(\bar{y}_2 - x_0 - x_1 \right),
\\
\rho_1(x;y_1,y_2) = \frac{1}{\bar{y}_2-y_1} \left[ \delta\left( x - \bar{y}_2 \right) - \delta\left(x - y_1 \right) \right],
\end{gather*}
where $\Theta(R)$ is equal to 1, where the relations $R$ are satisfied, and 0 elsewhere.


\section{Two-loop nonlogarithmic parts of the \boldmath $(x,y)$-correlators}\label{App:B}

The two-loop nonlogarithmic parts of the correlators read
\begin{align}
\tilde\Pi^\text{V}_{1,0}
=
{}&- \dot{W}_a \dot{H}_a + \frac12 W_a \ddot{H}_a + \sum_{\text{I}=a,\,b} \Biggl[ W_\text{I} \left( \frac12 \ddot{h}_\text{I} + \frac{\pi^2}3 \right) +\frac12 \ddot{W}_\text{I} - \dot{W}_\text{I} \dot{h}_\text{I} \Biggr]_+
\notag\\
&{}
 - \frac12 \hat{\mathbf{P}} \left[ \left( V_{(0)} \right)_{+(x)} \ddot{d} \right], \label{eq:PiV10}
\end{align}
\begin{align}
\tilde\Pi^\text{T}_{1,0}
={}&
\mathop{\hat{\mathbf{S}}} \biggl\{ \theta(x>y) \biggl[\frac12 \ln \left(\bar{z}\right) \ln (x \bar{x} y \bar{y}) + \frac14 \ln^2\left(\bar{z}\right) + \frac12 \ln \left(\bar{z}\right) + y \bar{x} + \frac{\pi^2}{12} \biggr] \biggr\}
\notag\\
&{}-\frac12\left( y \bar{x} + x \bar{y} \right) \mathop{\hat{\mathbf{S}}} \left[ \frac{\theta(x>y)}{x-y} \left( \frac{\pi^2}6 - \mathop{\mathrm{Li}_2}(z)\right) \right]
\notag\\
&{} + \Biggl[ W_b \left( \frac12 \ddot{h}_b + \frac{\pi^2}3 \right) +\frac12 \ddot{W}_b - \dot{W}_b \dot{h}_b - 2 \dot{W}_b + 2 W_b \dot{h}_b \Biggr]_+
 - \hat{\mathbf{P}} \left[ \left( V_b \right)_{+(x)} \left( \frac12 \ddot{d} - \dot{d} \right) \right]
\notag\\
&{}
-\delta(x-y) \left[ 2d - \frac54 \dot{d} + \frac18 \ddot{d} - \frac14 \ln(x) \ln(\bar{x}) d \right], \label{eq:PiT10}
\end{align}
\begin{eqnarray}
\tilde{\Pi}^\text{S}_{1,0}&=& \hat{\mathbf{P}} \left\{ \left( \hat{\mathbf{S}} \frac{\theta(y>x)}{y-x}\left[ 2 - \ln\left( x \bar{y} (y-x) \right) + \frac12 \ln^2\left( x \bar{y} (y-x) \right) + \frac{\pi^2}6 - \ln^2\left(y\bar{y}\right) \right]  \right)_{+(x)} \right\}
\notag\\
&& {} - \hat{\mathbf{P}} \left\{ \left( \hat{\mathbf{S}} \frac{\theta(y>x)}{y}\left[ 2 - \ln\left( x \bar{y} (y-x) \right) \right]  \right)_{+(x)} \right\}
- \hat{\mathbf{S}} \theta(y>x) \ln\left( x \bar{y} (y-x) \right) + 1
\notag\\
&& {} + 2 \ln\left( x\bar{x}y\bar{y} \right) - 2 \hat{\mathbf{S}} \left\{ \frac{\theta(x>y)}{x-y} \left[ \Li{2}(1) - \Li{2}(z) \right] \right\}
\notag\\
&& {} + \delta(x-y) \hat{\mathbf{S}} \left\{ \frac34 \ln^2(x) + \frac{x}2 \ln(x) - 10 \ln(x) + \frac{25}4 \right\}.
\label{eq:PiS10}
\end{eqnarray}

The functions $H_a$ and $h_\text{I}$, $\text{I}=a,b$ in eq.~\eqref{eq:PiV10} break the factorization of the two-loop correlator to one-loop subgraphs and are given explicitly by the following expressions, where $z=(y \bar x) /(x \bar y)$:
\begin{eqnarray}
	h_a
	&=&  h_a(x,y \vert \varepsilon )
	= h_a(y,x \vert \varepsilon )
\nonumber\\
	&=& \hat{\mathbf{S}}~\frac{\Theta(\bar z>0)}{4 } \biggl\{\frac{c^{-}_2(\varepsilon)}{(x-y)^{\varepsilon}}  \frac{\bar z}{z} \biggl[\frac{1}{x-y}-2\biggr]
+ \frac{c_3(\varepsilon)}{(x\bar{y})^\varepsilon}\biggl[2+\frac2{z}-\frac1{\bar{x}y} \biggr] \biggr\},
\\
	h_b &=& h_b(x, y \vert \varepsilon ) = h_b(y, x \vert \varepsilon )
	= \frac{\sin\left(\pi\varepsilon\right)}{\pi\varepsilon} \frac{1}{\lvert x-y \rvert^{\varepsilon}},
\\
H_a &=& H_a(x,y \vert \varepsilon )
 = H_a(y,x \vert \varepsilon )
	= \frac{\lvert x-y \rvert^{-\varepsilon}}{4}
		\left[ \frac{ g_{a1}(z \vert \varepsilon )}{\lvert x-y \rvert} +  g_{a2}(z \vert \varepsilon ) \right],
\end{eqnarray}
and
\begin{eqnarray}
g_{a1}(z \vert \varepsilon ) &=& \hat{\mathbf{S}}~ \Theta(\bar z>0) \frac{\bar z}{z} \biggl[c_2^+(\varepsilon) - c_3(\varepsilon) \bar{z}^\varepsilon + 2  \frac{1+z}{\bar z} \left(  \frac{z}{\bar z} \right)^{\varepsilon}
\nonumber \\
&&\hphantom{\frac{\bar z}{z} \biggl[c_2^+(\varepsilon) - c_3(\varepsilon) \bar{z}^\varepsilon + 2  \frac{1+z}{\bar z} }
\times \left[ I_{\bar z}\left(1+\varepsilon,-\varepsilon\right) - I_{\bar z}\left(1+2\varepsilon,-\varepsilon\right) \right] \biggr],
 \\
g_{a2}(z \vert \varepsilon ) &=& \hat{\mathbf{S}}~ 2\Theta(\bar z>0) \biggl[ -\varepsilon c_3(\varepsilon) \frac{1+z}{z} \bar{z}^{\varepsilon} - (1-\varepsilon) \left( \frac{\bar z}{z} \right)^{1-\varepsilon}
 \nonumber \\
&&\hphantom{\hat{\mathbf{S}} \biggl\{ \Theta(\bar z) \biggl[ -\varepsilon \mathrm{B}(\varepsilon,\varepsilon) \frac{1+z}{z} \bar{z}^{\varepsilon} -2}
\times\left[ I_{\bar z}\left(1+\varepsilon,-\varepsilon\right) - I_{\bar z}\left(1+2\varepsilon,-\varepsilon\right) \right] \biggr].
\end{eqnarray}
Here,
\begin{gather}
c_2^-(-\varepsilon) = c_2^+(\varepsilon) = \frac{1+\varepsilon}{\Gamma(1+\varepsilon)\Gamma(1-\varepsilon)},
\qquad
c_3(\varepsilon) = \frac{\Gamma(1+\varepsilon)}{\Gamma(1+2\varepsilon)\Gamma(1-\varepsilon)},
\\
I_{\bar z}\left(a,b\right) = \frac{\mathrm{B}_{\bar{z}}(a,b)}{\mathrm{B}(a,b)},
\end{gather}
and $\mathrm{B}_{\bar{z}}(a,b)$ is the incomplete beta function.
The corresponding Taylor series read
\begin{align}
	h_\text{I}(x,y \vert \varepsilon ) = 1 + \varepsilon \dot{h}_\text{I}(x,y) + \frac12 \varepsilon^2 \ddot{h}_\text{I}(x,y) + \dots,
\end{align}
\begin{align}
	H_a(x,y \vert \varepsilon ) &{}= H_a(x,y) + \varepsilon \dot{H}_a(x,y) + \frac12 \varepsilon^2 \ddot{H}_a(x,y) + \dots,
\\
	H_a(x,y) &{}= 0,
\\
	\dot{H}_a(x,y) &{}= \hat{\mathbf{S}} \frac{\Theta(y>x)}{2\bar{y}x} \left\{ \frac12 \left[1+\ln(\bar{x})-\ln\left(1-\frac{x}y \right) \right]- \bar{y}x-y\bar{x}\right\},
\\
	\ddot{H}_a(x,y) &{}= \hat{\mathbf{S}} \Biggl[ \frac{\Theta(x>y)}{z} \Biggl( \left\{ \frac{\bar{z} [\ln(\bar{z})-1]}{2(x-y)} +1+z \right\} \ln(x-y)
\notag\\&\hphantom{{}=\hat{\mathbf{S}} \Biggl[ \frac{\Theta(x>y)}{z} \Biggl( }  {}
	 -\frac{\bar{z}}{4 (x-y)} \ln^2(\bar{z}) -(1+z) \ln(\bar{z}) -\bar{z} \left[ \Li{2}(z) - \Li{2}(1) \right]
\notag\\&\hphantom{{}=\hat{\mathbf{S}} \Biggl[ \frac{\Theta(x>y)}{z} \Biggl(}
	{} + \left\{ \frac{\pi^2}{12} \bar{z} +(1+z) \left[ \Li{2}(z) - \Li{2}(1) \right] \right\} \frac{1}{x-y} \Biggr) \Biggr].
\end{align}


\section{\boldmath $(x,\underline{0})$-moments up to order $a_s^2 \beta_0$}\label{App:C}

Here, we write down the one-fold correlators as the following expansion:
\begin{gather}
\Pi^\text{X}(x;P^2) = F_\text{X} \frac{N_c}{\pi^2} \biggl\{ \tilde\Pi^\text{X}_{0}
			+ a_s C_F \tilde\Pi^\text{X}_{1}
			+ a_s^2 \cdot \left( \beta_0 C_F \tilde\Pi^\text{X}_{2[\beta]} + C_F\tilde\Pi^\text{X}_{2[0]} \right)
			+ \mathop{O}\left(a_s^3\right)
			\biggr\},
\\
\tilde\Pi^\text{X}_{i\bm{[a]}} = \tilde\Pi^\text{X}_{i\bm{[a]}}(x,P^2) = \sum_{k=0}^{i+1} \tilde\Pi^\text{X}_{i\bm{[a]},k}(x) \,  L^k ,
\end{gather}
where $F_\text{V} = \frac12$, $F_\text{T} = 1$, $F_\text{S} = \frac18$. The coefficients of the expansion are listed below:
\begin{align}
\tilde\Pi^\text{V}_{0,0} = -x\bar{x}\ln(x\bar{x}),
\qquad
\tilde\Pi^\text{V}_{0,1} = -x\bar{x},
\end{align}
\begin{subequations}
\begin{align}
\tilde\Pi^\text{V}_{1,0} ={}& x\bar{x} \left\{ 2 \mathop{\hat{\mathbf{S}}} \left[ \Li{3}(x) - \Li{2}(x) \ln(x) - \frac13 \ln^3 (x) \right] + \left( \frac{\pi^2}3 - 5 \right) \ln(x\bar{x}) + 9 - 16 \zeta_3 \right\}
\notag\\
{}& + \frac14 \mathop{\hat{\mathbf{S}}} \left[ x \ln^2 (x) \right],
\\
\tilde\Pi^\text{V}_{1,1} ={}& x\bar{x} \left[ \frac{\pi^2}3 - 5 - \ln^2\left(\frac{\bar{x}}{x}\right) \right],
\\
\tilde\Pi^\text{V}_{1,2} ={}& 0,
\end{align}
\end{subequations}
\begin{subequations}
\begin{align}
\tilde\Pi^\text{V}_{2[\beta],1} &{}= x \bar{x} \biggl\{ \mathop{\hat{\mathbf{S}}} \bigl[ 5 \Li{3}(x) - \Li{2}(x) \ln(x) \bigr] + \frac1{12} \ln^3 (x\bar{x}) - \frac14 \ln^2\left(\frac{\bar{x}}{x}\right) \ln(x\bar{x}) - \frac56 \ln^2\left(\frac{\bar{x}}{x}\right)
\notag\\
&\hphantom{{}= x \bar{x} \biggl\{} - \frac{\pi^2}3 \ln(x\bar{x}) + \frac5{18} \pi^2 - \frac76 \biggr\} + \mathop{\hat{\mathbf{S}}} \left\{ x \left[ \Li{2}(x) - \frac34 \ln^2(x) + \frac{31}{12} \ln(x) \right] \right\} - \frac{\pi^2}6, \\
\tilde\Pi^\text{V}_{2[\beta],2} &{}= -\frac12 \mathop{\hat{\mathbf{S}}} \left[ x \ln (x) \right],
\\
\tilde\Pi^\text{V}_{2[\beta],3} &{}= 0;
\end{align}
\end{subequations}

\begin{align}
\tilde\Pi^\text{T}_{0,0} = x\bar{x} \left[ 1+ \ln(x\bar{x}) \right],
\qquad
\tilde\Pi^\text{T}_{0,1} = x\bar{x},
\end{align}
\begin{subequations}
\begin{align}
\tilde\Pi^\text{T}_{1,0} ={}& x \bar{x} \biggl\{ 2 \mathop{\hat{\mathbf{S}}} \bigl[ -\Li{3}(x) + \ln(x) \Li{2}(x) \bigr] + \frac16 \ln^3 (x\bar{x}) + \frac12 \ln^2 \left(\frac{\bar{x}}{x}\right) \ln(x\bar{x}) + \frac18 \ln^2(x\bar{x})
\notag\\
&\hphantom{x \bar{x} \biggl\{} + \frac98 \ln^2 \left(\frac{\bar{x}}{x}\right) + \left( 6 - \frac{\pi^2}{3} \right) \ln (x\bar{x}) + 16 \zeta_3 - \frac{\pi^2}{3} - 8 \biggr\}
\notag\\
& - \mathop{\hat{\mathbf{S}}} \biggl[ x \ln(x) + \frac12 x \ln^2(x) \biggr],
\\
\tilde\Pi^\text{T}_{1,1} ={}& x \bar{x} \left[ 6 - \frac{\pi^2}{3} + \ln^2\left(\frac{\bar{x}}{x}\right) + \ln(x\bar{x}) \right],
\\
\tilde\Pi^\text{T}_{1,2} ={}& x \bar{x},
\end{align}
\end{subequations}
\begin{subequations}
\begin{align}
\tilde\Pi^\text{T}_{2[\beta],1} ={}& x \bar{x} \biggl\{ \mathop{\hat{\mathbf{S}}} \bigl[ -5 \Li{3}(x) + \Li{2}(x) \ln(x) \bigr] - \frac1{12} \ln^3 (x\bar{x}) + \frac14 \ln^2\left(\frac{\bar{x}}{x}\right) \ln(x\bar{x}) + \frac7{12} \ln^2\left(\frac{\bar{x}}{x}\right)
\notag\\
&\hphantom{ x \bar{x} \biggl\{}  -\frac14 \ln^2(x\bar{x}) + \frac13 \left( \frac{19}3 + \pi^2 \right) \ln(x\bar{x}) - \frac5{18} \pi^2 - \frac{193}{36} \biggr\}
\notag\\
&{} + \mathop{\hat{\mathbf{S}}} \left\{ x \left[ -2 \Li{2}(x) + \frac32 \ln^2(x) - \frac83 \ln(x) \right] \right\} + \frac{\pi^2}3,
\\
\tilde\Pi^\text{T}_{2[\beta],2} ={}& \frac12 x \bar{x} \left[ \frac{19}3 - \ln(x\bar{x}) \right] + \mathop{\hat{\mathbf{S}}} \bigl[ x \ln(x) \bigr],
\\
\tilde\Pi^\text{T}_{2[\beta],3} ={}& {-\frac13 x \bar{x}};
\end{align}
\end{subequations}
\begin{align}
\tilde\Pi^\text{S}_{0,0} = {-\ln(x\bar{x})},
\qquad
\tilde\Pi^\text{S}_{0,1} = {-1},
\end{align}
\begin{subequations}
\begin{align}
\tilde\Pi^\text{S}_{1,0} ={}& \mathop{\hat{\mathbf{S}}} \left[ 2 \Li{3}(x) - 2 \Li{2}(x) \ln(x) - \frac12 x \ln (x) \right] - \frac23 \ln^3(x\bar{x}) - \frac34 \ln^2(x\bar{x})
\notag\\
&{}+\ln(x\bar{x}) \left( 2 \ln(x) \ln(\bar{x}) - 10 + \frac{\pi^2}{3} \right) - \frac12 \ln(x) \ln(\bar{x}) -16 \zeta_3 - \frac{\pi^2}{3} + \frac{39}{2},
\\
\tilde\Pi^\text{S}_{1,1} ={}& \ln(x\bar{x}) - \ln^2\frac{x}{\bar{x}} + \frac{\pi^2}{3} - 15,
\\
\tilde\Pi^\text{S}_{1,2} ={}& 3,
\end{align}
\end{subequations}
\begin{subequations}
 \label{eq:PiS3}
\begin{align}
\tilde\Pi^\text{S}_{2[\beta],1} ={}& \mathop{\hat{\mathbf{S}}} \Bigl[ 5 \Li{3}(x) - \Li{2}(x) \ln(x) + x \ln (x) \Bigr] - \frac16 \ln^3(x\bar{x}) - \frac43 \ln^2(x\bar{x})
\notag\\
&{}+\ln(x\bar{x}) \left( \ln(x) \ln(\bar{x}) + \frac{31}{6} - \frac{\pi^2}{3} \right) + \frac{19}{3} \ln(x) \ln(\bar{x}) + \frac{11}{18} \pi^2 - \frac{479}{12}, \\
\tilde\Pi^\text{S}_{2[\beta],2} ={}& 10 - \frac12 \ln(x\bar{x}),
\\
\tilde\Pi^\text{S}_{2[\beta],3} ={}& {-1}.
\end{align}
\end{subequations}

The expansions above as well as rather cumbersome three-loop nonlogarithmic parts $\tilde\Pi^\text{X}_{2[\beta],0}$ of the moments are provided in an \texttt{.m} file appended to the \texttt{arXiv} version of this paper.


\section{Borel transform}\label{App:E}

The Borel transform with a parameter $\mu$ is defined as
\begin{align}
\hat{\mathbf{B}}_{(\mu)} \left[ f(t) \right] = \lim_{ t=n \mu,~ n \to \infty}
\frac{(-t)^n}{\Gamma(n)} \frac{d^n}{d t^n}f(t).
\end{align}
In this paper, we used the following special cases:
\begin{gather}
	\hat{\mathbf{B}}_{(\mu)}\left[ e^{-at} \right] = \delta(1-\mu a), \quad a >0,
\qquad\qquad
	\hat{\mathbf{B}}_{(\mu)} \left[ t^{-a} \right] = \frac{\mu^{-a}}{\Gamma(a)}, \quad a >0,
\\
\hat{\mathbf{B}}_{(\mu)} \left[ \ln^{m}(t) \right] = m(-)^m\, \frac{d^{m-1}}{d\varepsilon^{m-1}} \left[ \frac{e^{-\varepsilon l}}{\Gamma(1+\varepsilon)}\right]\Bigg|_{\varepsilon=0}
= - m \left[l_\text{B}^{m-1}- {m-1 \choose 2} \zeta_2 l_\text{B}^{m-2} + \ldots \right],  \label{eq:E-lnp}
\end{gather}
where $l=\ln\left(\mu\right)$ and  $l_\text{B}=\ln(\mu e^{-\gamma_\text{E}})$.
For the case of scalar-scalar correlator, one has to Borel transform the terms proportional to  $P^2 \ln\left(P^2/\mu^2 \right)^k$, which can be done with the help of eq.~(\ref{eq:E-lnp}) and the relation below,
\begin{equation} \label{eq:E-plnp}
\hat{\mathbf{B}}_{(\mu)} \left[p \ln^{n}(p) \right]=-\mu \left\{\hat{\mathbf{B}}_{(\mu)}[\ln^{n}(p)]+n \hat{\mathbf{B}}_{(\mu)}[\ln^{n-1}(p)]\right\}.
\end{equation}
In particular,
\begin{subequations}
 \label{eq:Borel-pLp}
 \begin{gather}
\hat{\mathbf{B}}_{(\mu)} \left[ p\ln(p) \right] = \mu,
\qquad\qquad
\hat{\mathbf{B}}_{(\mu)} \left[ p\ln^2(p) \right] = 2\mu \left(l_\text{B} +1\right),
\\
\hat{\mathbf{B}}_{(\mu)} \left[ p\ln^3(p) \right] = 3\mu \left(l^2_\text{B} + 2l_\text{B} -\frac{\pi^2}{6}\right).
\end{gather}
 \end{subequations}


\section{Distribution amplitudes of twist 2 and 3 for \boldmath $\pi$ and $\rho$ mesons}
 \label{App:DA}
Distribution amplitudes (DA) of hadrons appear as a result of applying factorization theorems
to hard exclusive processes with hadrons,
they describe the parton degrees of freedom in the soft hadron part of the factorized amplitudes.
The DAs parameterize, in the collinear direction, the matrix elements of the gauge invariant nonlocal operators sandwiched between the vacuum and the hadron state.
The DAs are ordered by their increasing twist.
Indeed, the two particle DAs presented below describe the partition
of longitudinal-momentum fractions between the valence quark, $x$, and antiquark, $1-x$.
 The twist-2 DA
$\varphi_\pi(x,\mu^2)$ for the pion and $\varphi_\rho^\text{L}(x,\mu^2)$
for the longitudinal $\rho$ meson, are defined as
\begin{align}
\langle 0\rvert \bar d(0) \gamma_\nu \gamma_5\,[0,z] u(z)\lvert \pi^{+}(p) \rangle \Big|_{z^2=0}&= i f_\pi p_\nu
 \int_{0}^{1} dx \, e^{-ix (z \cdot p)}\,
  \varphi_\pi(x,\mu^2),
  \quad
  \int_{0}^{1} dx \,  \varphi_\pi(x,\mu^2)=1; \label{def:pi}
\\
 \langle 0\rvert \bar d(0) \gamma_\nu\,[0,z] u(z)\lvert \rho(p,\lambda) \rangle \Big|_{z^2=0}
&=
  f_\rho p_\nu\,
   \int_{0}^{1} dx\, e^{-ix (z \cdot p)}\,   \varphi_\rho^\text{L}(x,\mu^2),
  \quad
  \int_{0}^{1} dx \,  \varphi_\rho^\text{L}(x,\mu^2)=1,
\label{def:rhoL}
\end{align}
where $p_{\nu}$ and $\mu^2 $ are the meson momentum and the factorization scale $(\mu^2=\mu^2_\text{F})$.
The path-ordered gauge link $[0,z]$ with the integration along the straight line on the light cone is
\begin{equation}
[0,z]=\text{P}\exp\left(-ig\int_{0}^{z}A_\nu(t)\,dt^\nu \right)\,.
\end{equation}
On the other hand, the transverse $\rho$-meson DA,
$\varphi_\rho^\text{T}(x,\mu^2)$, is given by
\begin{eqnarray}
  \langle 0\vert \bar d(0) \sigma_{\mu\nu}\,[0,z]u(z)\vert\rho(p,\lambda) \rangle \Big|_{z^2=0}
=
  i f_\rho^\text{T}(\varepsilon^{(\lambda)}_\mu p_\nu-\varepsilon^{(\lambda)}_\nu p_\mu)\,
  \int_{0}^{1} dx\ e^{-ix (z \cdot p)}
  \varphi_\rho^\text{T}(x,\mu^2) ,
\label{def:rhoT}
\end{eqnarray}
where $\varepsilon^{(\lambda)}_\mu$ is the polarization vector of the $\rho$  meson,  $\lambda$ --- its helicity. \\
Below we neglect the contributions of twist-3 three-particle DA, see \cite{Ball:1998je,Ball:2006wn} and eqs.~(20,21) in \cite{Braun:1988qv}:
\begin{gather}
\langle 0|\bar{q}(0) i\gamma_5\,[0,z] q(z)|\pi(p)\rangle|_{z^2=0} =  \frac{f_\pi m^2_\pi}{(m_d+m_u)}\int^{1}_{0} dx e^{-ipzx}\varphi^p_{3;\pi}(x,\mu^2),
\\
\int^{1}_{0}\varphi^p_{3;\pi}(x,\mu^2) dx=1, \qquad\qquad \varphi^{p,~as}_{3;\pi}(x)=1,
\end{gather}
\begin{gather}
\langle 0|\bar{q}(0) i\sigma_{\mu \alpha}z^\alpha\gamma_5\,[0,z] q(z)|\pi(p)\rangle|_{z^2=0} = -\frac{i}{6}(pz)z_\mu\frac{f_\pi m^2_\pi}{(m_d+m_u)}\int^{1}_{0} dx e^{-ipzx}\varphi^\sigma_{3;\pi}(x,\mu^2), \\
\int^{1}_{0}\varphi^{\sigma}_{3;\pi}(x,\mu^2) dx=1, \qquad\qquad \varphi^{\sigma,~as}_{3;\pi}(x)=6x\bar{x}\,.
\end{gather}

\newcommand{\noopsort}[1]{} \newcommand{\printfirst}[2]{#1}
  \newcommand{\singleletter}[1]{#1} \newcommand{\switchargs}[2]{#2#1}

\providecommand{\href}[2]{#2}\begingroup\raggedright\endgroup


\begin{thebibliography}{10}

\bibitem{Craigie:1983fb}
N.~S. Craigie, V.~K. Dobrev and I.~T. Todorov, \emph{{Conformally covariant
  composite operators in quantum chromodynamics}},
  \href{https://doi.org/10.1016/0003-4916(85)90118-6}{\emph{Annals Phys.}
  {\bfseries 159} (1985) 411}.

\bibitem{Gracey:2009da}
J.~A. Gracey, \emph{{Three loop $\overline{\mbox{MS}}$ operator correlation
  functions for deep inelastic scattering in the chiral limit}},
  \href{https://doi.org/10.1088/1126-6708/2009/04/127}{\emph{J. High Energ.
  Phys.} {\bfseries 04} (2009) 127}
  [\href{https://arxiv.org/abs/0903.4623}{{\ttfamily 0903.4623}}].

\bibitem{Radyushkin:1983wh}
A.~V. Radyushkin, \emph{{On spectral properties of parton correlation functions
  and multiparton wave functions}},
  \href{https://doi.org/10.1016/0370-2693(83)91116-4}{\emph{Phys. Lett. B}
  {\bfseries 131} (1983) 179}.

\bibitem{Mikhailov:1984ii}
S.~V. Mikhailov and A.~V. Radyushkin, \emph{{Evolution kernels in QCD: Two-loop
  calculation in Feynman gauge}},
  \href{https://doi.org/10.1016/0550-3213(85)90213-5}{\emph{Nucl. Phys. B}
  {\bfseries 254} (1985) 89}.

\bibitem{Mikhailov:2018udp}
S.~V. Mikhailov and N.~Volchanskiy, \emph{{Two-loop kite master integral for a
  correlator of two composite vertices}},
  \href{https://doi.org/10.1007/JHEP01(2019)202}{\emph{J. High Energ. Phys.}
  {\bfseries 01} (2019) 202}
  [\href{https://arxiv.org/abs/1812.02164}{{\ttfamily 1812.02164}}].

\bibitem{Chetyrkin:2010dx}
K.~G. Chetyrkin and A.~Maier, \emph{{Massless correlators of vector, scalar and
  tensor currents in position space at orders $\alpha_s^3$ and $\alpha_s^4$:
  Explicit analytical results}},
  \href{https://doi.org/10.1016/j.nuclphysb.2010.11.007}{\emph{Nucl. Phys. B}
  {\bfseries 844} (2011) 266}
  [\href{https://arxiv.org/abs/1010.1145}{{\ttfamily 1010.1145}}].

\bibitem{Mikhailov:1988nz-JINRrep}
S.~V. Mikhailov and A.~V. Radyushkin, \emph{{Quark Condensate Nonlocality and
  Pion Wave Function in QCD: General Formalism}}, {\emph{preprint
  JINR-P2-88-103} (1988) }
  [\href{https://arxiv.org/abs/http://inspirehep.net/record/262441/files/JINR-P2-88-103.pdf}{{\ttfamily
  http://inspirehep.net/record/262441/files/JINR-P2-88-103.pdf}}].

\bibitem{Mikhailov:1997zg}
S.~V. Mikhailov, \emph{{Renormalon chains contributions to the nonsinglet
  evolution kernels in $\left[\phi^3\right]_6$ in six-dimensions and QCD}},
  \href{https://doi.org/10.1016/S0370-2693(97)01198-2}{\emph{Phys. Lett. B}
  {\bfseries 416} (1998) 421}
  [\href{https://arxiv.org/abs/hep-ph/9706326}{{\ttfamily hep-ph/9706326}}].

\bibitem{Mikhailov:1998xi}
S.~V. Mikhailov, \emph{{Renormalon chains contributions to nonsinglet
  evolutional kernels in QCD}},
  \href{https://doi.org/10.1016/S0370-2693(98)00585-1}{\emph{Phys. Lett. B}
  {\bfseries 431} (1998) 387}
  [\href{https://arxiv.org/abs/hep-ph/9804263}{{\ttfamily hep-ph/9804263}}].

\bibitem{Mikhailov:1985cm}
S.~V. Mikhailov and A.~V. Radyushkin, \emph{{Structure of two-loop evolution
  kernels and evolution of the pion wave function in $\phi_{(6)}^3$ and QCD}},
  \href{https://doi.org/10.1016/0550-3213(86)90248-8}{\emph{Nucl. Phys. B}
  {\bfseries 273} (1986) 297}.

\bibitem{Ball:1996tb}
P.~Ball and V.~M. Braun, \emph{{The Rho meson light cone distribution
  amplitudes of leading twist revisited}},
  \href{https://doi.org/10.1103/PhysRevD.54.2182}{\emph{Phys. Rev.} {\bfseries
  D54} (1996) 2182} [\href{https://arxiv.org/abs/hep-ph/9602323}{{\ttfamily
  hep-ph/9602323}}].

\bibitem{Mikhailov:1988nz}
S.~V. Mikhailov and A.~V. Radyushkin, \emph{{Quark Condensate Nonlocality and
  Pion Wave Function in QCD: General Formalism}}, {\emph{Sov. J. Nucl. Phys.}
  {\bfseries 49} (1989) 494}.

\bibitem{Mikhailov:2008my}
S.~V. Mikhailov and A.~A. Vladimirov, \emph{{ERBL and DGLAP kernels for
  transversity distributions. Two-loop calculations in covariant gauge}},
  \href{https://doi.org/10.1016/j.physletb.2008.11.051}{\emph{Phys. Lett. B}
  {\bfseries 671} (2009) 111}
  [\href{https://arxiv.org/abs/0810.1647}{{\ttfamily 0810.1647}}].

\bibitem{Braun2017}
V.~M. Braun, A.~N. Manashov, S.~Moch and M.~Strohmaier, \emph{Three-loop
  evolution equation for flavor-nonsinglet operators in off-forward
  kinematics}, \href{https://doi.org/10.1007/JHEP06(2017)037}{\emph{J. High
  Energy Phys.} {\bfseries 2017} (2017) 37}
  [\href{https://arxiv.org/abs/1703.09532}{{\ttfamily 1703.09532}}].

\bibitem{Ball:1998je}
P.~Ball, \emph{{Theoretical update of pseudoscalar meson distribution
  amplitudes of higher twist: the nonsinglet case}},
  \href{https://doi.org/10.1088/1126-6708/1999/01/010}{\emph{J. High Energ.
  Phys.} {\bfseries 01} (1999) 010}
  [\href{https://arxiv.org/abs/hep-ph/9812375}{{\ttfamily hep-ph/9812375}}].

\bibitem{Ball:2006wn}
P.~Ball, V.~M. Braun and A.~Lenz, \emph{{Higher-twist distribution amplitudes
  of the K meson in QCD}},
  \href{https://doi.org/10.1088/1126-6708/2006/05/004}{\emph{J. High Energ.
  Phys.} {\bfseries 05} (2006) 004}
  [\href{https://arxiv.org/abs/hep-ph/0603063}{{\ttfamily hep-ph/0603063}}].

\bibitem{Kataev:2016aib}
A.~L. Kataev and S.~V. Mikhailov, \emph{{The $\{\beta\}$-expansion formalism in
  perturbative QCD and its extension}},
  \href{https://doi.org/10.1007/JHEP11(2016)079}{\emph{J. High Energ. Phys.}
  {\bfseries 11} (2016) 079}
  [\href{https://arxiv.org/abs/1607.08698}{{\ttfamily 1607.08698}}].

\bibitem{Kotikov:2019bqo}
A.~V. {Kotikov} and S.~{Teber}, \emph{{Landau-Khalatnikov-Fradkin
  transformation and the mystery of even $\zeta$-values in Euclidean massless
  correlators}}, \href{https://doi.org/10.1103/PhysRevD.100.105017}{\emph{Phys.
  Rev. D} {\bfseries 100} (2019) 105017}
  [\href{https://arxiv.org/abs/1906.10930}{{\ttfamily 1906.10930}}].

\bibitem{Baikov:2019zmy}
P.~A. Baikov and K.~G. Chetyrkin, \emph{{Transcendental structure of multiloop
  massless correlators and anomalous dimensions}},
  \href{https://doi.org/10.1007/JHEP10(2019)190}{\emph{J. High Energ. Phys.}
  {\bfseries 10} (2019) 190}
  [\href{https://arxiv.org/abs/1908.03012}{{\ttfamily 1908.03012}}].

\bibitem{Mikhailov:2020izb}
S.~V. Mikhailov and N.~Volchanskiy, \emph{{Radiative corrections to QCD SR for
  meson distribution amplitudes up to $O(\alpha_s^2\beta_0)$}},
  \href{https://doi.org/10.1088/1742-6596/1435/1/012059}{\emph{J. Phys. Conf.
  Ser.} {\bfseries 1435} (2020) 012059}.

\bibitem{Brodsky:1982gc}
S.~J. Brodsky, G.~P. Lepage and P.~B. Mackenzie, \emph{{On the elimination of
  scale ambiguities in perturbative quantum chromodynamics}},
  \href{https://doi.org/10.1103/PhysRevD.28.228}{\emph{Phys. Rev. D} {\bfseries
  28} (1983) 228}.

\bibitem{Pimikov:2013usa}
A.~V. Pimikov, S.~V. Mikhailov and N.~G. Stefanis, \emph{{Rho meson
  distribution amplitudes from QCD sum rules with nonlocal condensates}},
  \href{https://doi.org/10.1007/s00601-014-0815-5}{\emph{Few Body Syst.}
  {\bfseries 55} (2014) 401} [\href{https://arxiv.org/abs/1312.2776}{{\ttfamily
  1312.2776}}].

\bibitem{Stefanis:2015qha}
N.~Stefanis and A.~Pimikov, \emph{{Chimera distribution amplitudes for the pion
  and the longitudinally polarized \ensuremath{\rho}-meson}},
  \href{https://doi.org/10.1016/j.nuclphysa.2015.11.002}{\emph{Nucl. Phys. A}
  {\bfseries 945} (2016) 248}
  [\href{https://arxiv.org/abs/1506.01302}{{\ttfamily 1506.01302}}].

\bibitem{Bali:2019dqc}
G.~S. {Bali}, V.~M. {Braun}, S.~{B{\"u}rger}, M.~{G{\"o}ckeler}, M.~{Gruber},
  F.~{Hutzler} et~al., \emph{{Light-cone distribution amplitudes of
  pseudoscalar mesons from lattice QCD}},
  \href{https://doi.org/10.1007/JHEP08(2019)065}{\emph{J. High Energ. Phys.}
  {\bfseries 08} (2019) 065}
  [\href{https://arxiv.org/abs/1903.08038}{{\ttfamily 1903.08038}}].

\bibitem{Braun:2016wnx}
V.~M. {Braun}, P.~C. {Bruns}, S.~{Collins}, J.~A. {Gracey}, M.~{Gruber},
  M.~{G{\"o}ckeler} et~al., \emph{{The $\rho$-meson light-cone distribution
  amplitudes from lattice QCD}},
  \href{https://doi.org/10.1007/JHEP04(2017)082}{\emph{J. High Energ. Phys.}
  {\bfseries 04} (2017) 082}
  [\href{https://arxiv.org/abs/1612.02955}{{\ttfamily 1612.02955}}].

\bibitem{Braun:1988qv}
V.~M. Braun and I.~E. Filyanov, \emph{{QCD sum rules in exclusive kinematics
  and pion wave function}}, \href{https://doi.org/10.1007/BF01548594}{\emph{Z.
  Phys. C} {\bfseries 44} (1989) 157}.

\end{thebibliography}
\end{document}